\documentclass[a4paper,11pt]{article}

\usepackage[top=1.5in,right=1.0in,left=1.0in,bottom=1.0in]{geometry}
 
\usepackage{graphics,graphicx}
\usepackage{subfig}
\usepackage{float}
\usepackage{booktabs}
\usepackage{longtable}
\usepackage{supertabular}
\usepackage{tabularx}
\usepackage{multirow}
\usepackage{array}
\usepackage{tabu}
\usepackage{mathtools}
\usepackage{hyperref}

\newcommand{\otoprule}{\midrule[\heavyrulewidth]} 

\usepackage{amsmath}
\usepackage{amsfonts}
\usepackage{amssymb}
\usepackage{siunitx} 

\usepackage{enumerate}
\usepackage{sectsty}
\usepackage[affil-it,auth-sc]{authblk}
\usepackage{color}
\usepackage[dvipsnames]{xcolor}

\sectionfont{\normalsize}
\subsectionfont{\normalsize}

\usepackage{tikz}
\usetikzlibrary{arrows,decorations,backgrounds,shapes, snakes}
\usetikzlibrary{positioning}
\usetikzlibrary{matrix} 
\usepackage{varwidth}
\usepackage[most]{tcolorbox}
\usetikzlibrary{shapes.geometric,arrows.meta,decorations.markings}
\usepackage{fancyhdr}            
\fancypagestyle{plain}{%
  \fancyhead{}                                     
  \fancyfoot[CE,CO]{\thepage}       
  \fancyhead[CE,CO]{\textcolor[rgb]{0.64,0.15,0.15}{Published in \emph{Journal of the Mechanics and Physics of Solids} (2019) \textbf{130}, 82--100
  \\
doi: https://doi.org/10.1016/j.jmps.2019.05.009}}
\setlength{\headheight}{14.5pt}}

\begin{document}

\newcommand{\singlespace}{\baselineskip=12pt\lineskiplimit=0pt\lineskip=0pt}
\def\ds{\displaystyle}

\tikzstyle{every picture}+=[remember picture]

\newcommand{\beq}{\begin{equation}}
\newcommand{\eeq}{\end{equation}}
\newcommand{\lb}{\label}
\newcommand{\ph}{\phantom}
\newcommand{\beqar}{\begin{eqnarray}}
\newcommand{\eeqar}{\end{eqnarray}}
\newcommand{\barr}{\begin{array}}
\newcommand{\earr}{\end{array}}
\newcommand{\jump}{\parallel}
\newcommand{\Ehat}{\hat{E}}
\newcommand{\That}{\hat{\bf T}}
\newcommand{\Ahat}{\hat{A}}
\newcommand{\chat}{\hat{c}}
\newcommand{\shat}{\hat{s}}
\newcommand{\khat}{\hat{k}}
\newcommand{\muhat}{\hat{\mu}}
\newcommand{\mc}{M^{\scriptscriptstyle C}}
\newcommand{\mei}{M^{\scriptscriptstyle M,EI}}
\newcommand{\mec}{M^{\scriptscriptstyle M,EC}}
\newcommand{\hbeta}{{\hat{\beta}}}
\newcommand{\rec}[2]{\left( #1 #2 \ds{\frac{1}{#1}}\right)}
\newcommand{\rep}[2]{\left( {#1}^2 #2 \ds{\frac{1}{{#1}^2}}\right)}
\newcommand{\derp}[2]{\ds{\frac {\partial #1}{\partial #2}}}
\newcommand{\derpn}[3]{\ds{\frac {\partial^{#3}#1}{\partial #2^{#3}}}}
\newcommand{\dert}[2]{\ds{\frac {d #1}{d #2}}}
\newcommand{\dertn}[3]{\ds{\frac {d^{#3} #1}{d #2^{#3}}}}
\newcommand{\ct}{\captionof{table}}
\newcommand{\cf}{\captionof{figure}}

\def\c{{\circ}}
\def\bob{{\, \underline{\overline{\otimes}} \,}}
\def\ob{{\, \underline{\otimes} \,}}
\def\scalp{\mbox{\boldmath$\, \cdot \, $}}
\def\gdp{\makebox{\raisebox{-.215ex}{$\Box$}\hspace{-.778em}$\times$}}
\def\daa{\makebox{\raisebox{-.050ex}{$-$}\hspace{-.550em}$: ~$}}
\def\mK{\mbox{${\mathcal{K}}$}}
\def\cK{\mbox{${\mathbb {K}}$}}

\def\Xint#1{\mathchoice
   {\XXint\displaystyle\textstyle{#1}}%
   {\XXint\textstyle\scriptstyle{#1}}%
   {\XXint\scriptstyle\scriptscriptstyle{#1}}%
   {\XXint\scriptscriptstyle\scriptscriptstyle{#1}}%
   \!\int}
\def\XXint#1#2#3{{\setbox0=\hbox{$#1{#2#3}{\int}$}
     \vcenter{\hbox{$#2#3$}}\kern-.5\wd0}}
\def\ddashint{\Xint=}
\def\fpint{\Xint=}
\def\dashint{\Xint-}
\def\cpvint{\Xint-}
\def\intl{\int\limits}
\def\cpvintl{\cpvint\limits}
\def\fpintl{\fpint\limits}
\def\ointl{\oint\limits}
\def\bA{{\bf A}}
\def\ba{{\bf a}}
\def\bB{{\bf B}}
\def\bb{{\bf b}}
\def\bc{{\bf c}}
\def\bC{{\bf C}}
\def\bD{{\bf D}}
\def\bE{{\bf E}}
\def\be{{\bf e}}
\def\bbf{{\bf f}}
\def\bF{{\bf F}}
\def\bG{{\bf G}}
\def\bg{{\bf g}}
\def\bi{{\bf i}}
\def\bH{{\bf H}}
\def\bK{{\bf K}}
\def\bL{{\bf L}}
\def\bM{{\bf M}}
\def\bN{{\bf N}}
\def\bn{{\bf n}}
\def\bm{{\bf m}}
\def\b0{{\bf 0}}
\def\bo{{\bf o}}
\def\bX{{\bf X}}
\def\bx{{\bf x}}
\def\bP{{\bf P}}
\def\bp{{\bf p}}
\def\bQ{{\bf Q}}
\def\bq{{\bf q}}
\def\bR{{\bf R}}
\def\bS{{\bf S}}
\def\bs{{\bf s}}
\def\bT{{\bf T}}
\def\bt{{\bf t}}
\def\bU{{\bf U}}
\def\bu{{\bf u}}
\def\bv{{\bf v}}
\def\bw{{\bf w}}
\def\bW{{\bf W}}
\def\by{{\bf y}}
\def\bz{{\bf z}}
\def\T{{\bf T}}
\def\Te{\textrm{T}}
\def\Id{{\bf I}}
\def\bxi{\mbox{\boldmath${\xi}$}}
\def\balpha{\mbox{\boldmath${\alpha}$}}
\def\bbeta{\mbox{\boldmath${\beta}$}}
\def\bepsilon{\mbox{\boldmath${\epsilon}$}}
\def\bvarepsilon{\mbox{\boldmath${\varepsilon}$}}
\def\bomega{\mbox{\boldmath${\omega}$}}
\def\bphi{\mbox{\boldmath${\phi}$}}
\def\bsigma{\mbox{\boldmath${\sigma}$}}
\def\bfeta{\mbox{\boldmath${\eta}$}}
\def\bDelta{\mbox{\boldmath${\Delta}$}}
\def\btau{\mbox{\boldmath $\tau$}}
\def\tr{{\rm tr}}
\def\dev{{\rm dev}}
\def\div{{\rm div}}
\def\Div{{\rm Div}}
\def\Grad{{\rm Grad}}
\def\grad{{\rm grad}}
\def\Lin{{\rm Lin}}
\def\Sym{{\rm Sym}}
\def\Skw{{\rm Skew}}
\def\abs{{\rm abs}}
\def\Re{{\rm Re}}
\def\Im{{\rm Im}}
\def\capB{\mbox{\boldmath${\mathsf B}$}}
\def\capC{\mbox{\boldmath${\mathsf C}$}}
\def\capD{\mbox{\boldmath${\mathsf D}$}}
\def\capE{\mbox{\boldmath${\mathsf E}$}}
\def\capG{\mbox{\boldmath${\mathsf G}$}}
\def\tcapG{\tilde{\capG}}
\def\capH{\mbox{\boldmath${\mathsf H}$}}
\def\capK{\mbox{\boldmath${\mathsf K}$}}
\def\capL{\mbox{\boldmath${\mathsf L}$}}
\def\capM{\mbox{\boldmath${\mathsf M}$}}
\def\capR{\mbox{\boldmath${\mathsf R}$}}
\def\capW{\mbox{\boldmath${\mathsf W}$}}

\def\i{\mbox{${\mathrm i}$}}
\def\mC{\mbox{\boldmath${\mathcal C}$}}
\def\mB{\mbox{${\mathcal B}$}}
\def\mE{\mbox{${\mathcal{E}}$}}
\def\mL{\mbox{${\mathcal{L}}$}}
\def\mK{\mbox{${\mathcal{K}}$}}
\def\mV{\mbox{${\mathcal{V}}$}}
\def\C{\mbox{\boldmath${\mathcal C}$}}
\def\E{\mbox{\boldmath${\mathcal E}$}}

\def\AAM{{\it Advances in Applied Mechanics }}
\def\ACME{{\it Arch. Comput. Meth. Engng.}}
\def\ARMA{{\it Arch. Rat. Mech. Analysis}}
\def\AMR{{\it Appl. Mech. Rev.}}
\def\ASCEEM{{\it ASCE J. Eng. Mech.}}
\def\ACTA{{\it Acta Mater.}}
\def\CMAME {{\it Comput. Meth. Appl. Mech. Engrg.}}
\def\CRAS{{\it C. R. Acad. Sci. Paris}}
\def\CRM{{\it Comptes Rendus M\'ecanique}}
\def\EFM{{\it Eng. Fracture Mechanics}}
\def\EJMA{{\it Eur.~J.~Mechanics-A/Solids}}
\def\IJES{{\it Int. J. Eng. Sci.}}
\def\IJF{{\it Int. J. Fracture}}
\def\IJMS{{\it Int. J. Mech. Sci.}}
\def\IJNAMG{{\it Int. J. Numer. Anal. Meth. Geomech.}}
\def\IJP{{\it Int. J. Plasticity}}
\def\IJSS{{\it Int. J. Solids Structures}}
\def\IngA{{\it Ing. Archiv}}
\def\JAM{{\it J. Appl. Mech.}}
\def\JAP{{\it J. Appl. Phys.}}
\def\JAE{{\it J. Aerospace Eng.}}
\def\JE{{\it J. Elasticity}}
\def\JM{{\it J. de M\'ecanique}}
\def\JMPS{{\it J. Mech. Phys. Solids}}
\def\JSV{{\it J. Sound and Vibration}}
\def\MACRO{{\it Macromolecules}}
\def\MMT{{\it Mech. Mach. Th.}}
\def\MOM{{\it Mech. Materials}}
\def\MMS{{\it Math. Mech. Solids}}
\def\MMT{{\it Metall. Mater. Trans. A}}
\def\MPCPS{{\it Math. Proc. Camb. Phil. Soc.}}
\def\MSE{{\it Mater. Sci. Eng.}}
\def\NATURE{{\it Nature}}
\def\NATUREM{{\it Nature Mater.}}
\def\PHIL{{\it Phil. Trans. R. Soc.}}
\def\PMPS{{\it Proc. Math. Phys. Soc.}}
\def\PNAS{{\it Proc. Nat. Acad. Sci.}}
\def\PRE{{\it Phys. Rev. E}}
\def\PRL{{\it Phys. Rev. Letters}}
\def\PRSL{{\it Proc. R. Soc.}}
\def\RIIT{{\it Rozprawy Inzynierskie - Engineering Transactions}}
\def\ROCK{{\it Rock Mech. and Rock Eng.}}
\def\QAM{{\it Quart. Appl. Math.}}
\def\QJMAM{{\it Quart. J. Mech. Appl. Math.}}
\def\SCIENCE{{\it Science}}
\def\SCRMAT{{\it Scripta Mater.}}
\def\SM{{\it Scripta Metall.}}
\def\ZAMM{{\it Z. Angew. Math. Mech.}}
\def\ZAMP{{\it Z. Angew. Math. Phys.}}
\def\ZVDI{{\it Z. Verein. Deut. Ing.}}

\def\salto#1#2{
[\mbox{\hspace{-#1em}}[#2]\mbox{\hspace{-#1em}}]}

\renewcommand\Affilfont{\itshape}
\setlength{\affilsep}{1em}
\renewcommand\Authsep{, }
\renewcommand\Authand{ and }
\renewcommand\Authands{ and }
\setcounter{Maxaffil}{2}

\title{Configurational forces and nonlinear structural dynamics}

\author{C. Armanini}
\author{F. Dal Corso}
\author{D. Misseroni}
\author{D. Bigoni}
\affil[]{DICAM, University of Trento, via~Mesiano~77, I-38123 Trento, Italy.

Corresponding author: bigoni@ing.unitn.it
}

\date{}
\maketitle

\begin{abstract}

Configurational, or Eshelby-like, forces are shown to strongly influence the nonlinear dynamics of an elastic rod constrained with a frictionless sliding sleeve at one end and 
with an attached mass at the other end. 
The configurational force, generated at the sliding sleeve constraint and proportional  to the square of the bending moment realized there, has been 
so far investigated only under quasi-static setting and is now confirmed (through a variational argument) to be present within a dynamic framework.  The deep influence of configurational forces on the dynamics is shown both theoretically (through the development 
of a dynamic nonlinear model in which 
the rod is treated as a nonlinear spring, obeying the Euler elastica, with negligible inertia) 
and experimentally (through a specifically designed experimental set-up). 
During the nonlinear dynamics, the elastic rod may slip alternatively in and out from the sliding sleeve,  becoming a sort of nonlinear oscillator 
displaying a motion  eventually ending with the rod completely injected into or completely ejected from the sleeve. 
The present results may find applications in the dynamics of compliant and extensible devices, for instance, to guide the movement of a retractable and flexible robot arm.

\end{abstract}


\section{Introduction}\lb{INTRO}

Nonlinear structural dynamics breaks the limits of traditional linear elastic design, to create elements
working much beyond the realm of linearized kinematics, fully inside the nonlinear range, so matching the strong requirements 
imposed by soft robotics \cite{gravagne, renda, zheng}, flexible locomotion devices \cite{cazzolli,tsuda,wang}, metastructures \cite{ada, nadkarni, sugino}, architected structures for vibration mitigation \cite{carta, garau, nievesJMPS},  and  morphable structures \cite{gomez,koch,pandey}.

Within this context, the influence of configurational (or \lq Eshelby-like') forces on the nonlinear dynamics of structures is investigated. Configurational forces, introduced in solid mechanics by Eshelby \cite{eshelby1,eshelby2,eshelby3, eshelby4} to model interactions between dislocations or forces driving crack propagation, have been recently shown to be possible in structural mechanics too  \cite{bigonieshelby}. The action  of configurational forces on structures have been exploited to provide unexpected quasi-static response \cite{bigoniblade, bosiarmscale,bosidripping} and propulsion \cite{bigonitorsionalgun, dalcorsosnake},  have been explained through a material force balance \cite{hanna,oreilly1,oreilly2,singh} and have been used to investigate constrained buckling problems \cite{liakou1,liakou2,liakou3}. It is also worth noting that the recent research on configurational forces in structural mechanics has eventually  
inspired a new interpretation of their action in solids \cite{ballarini}.

\begin{figure}[!h]
\renewcommand{\figurename}{\footnotesize{Fig.}}
    \begin{center}
   \includegraphics[width=1\textwidth]{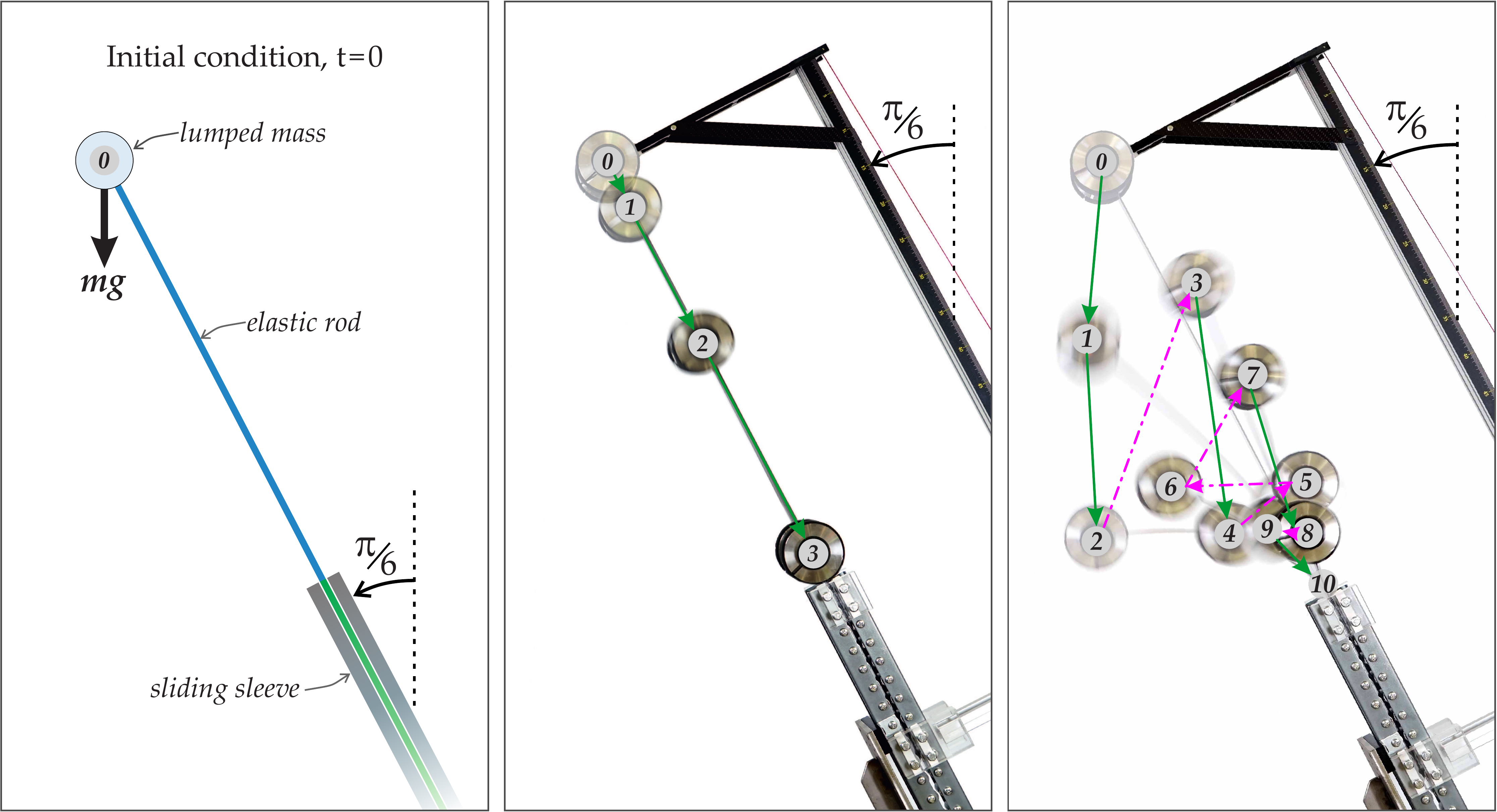}
    \caption{\footnotesize 
		Stroboscopic photos  taken after the release from the undeformed rest configuration (left)  of a highly stiff (center) and elastic (right) rod,  having an attached mass at one end and constrained by a frictionless sliding sleeve at the other. 	
When the rod is highly stiff (center), the system simply slides rectilinearly down along the sliding sleeve direction. When the rod is compliant (right), a complex motion is realized, in which injection into the sliding sleeve alternates with ejection because of the action of a configurational force (generated at the sliding sleeve end), strongly affecting the dynamics. While the rigid-body injection finishes in 0.39 sec (snapshot (3) in the center),  the compliant rod is completely injected into the 
sliding sleeve in 10.55 sec (snapshot not reported for clarity). 
Green (magenta) arrows connecting two successive snapshots denote the injective (ejective) motion occurring between them, so that from (0) to (2) the compliant rod is slipping inside the sleeve, from (2) to (3) it is ejected and so on. The timeframe of the snapshots for each of the two systems is reported in Table \ref{tabella1}.}
    \label{intro}
    \end{center}
\end{figure}

With the aim of investigating the role of configurational forces developing during the  motion of an elastic structure, 
the dynamic problem of a rod with a concentrated mass  attached at one end and subject to gravity is analyzed when partially inserted into a frictionless sliding 
sleeve at the other, Fig. \ref{intro} (left).  
Initially, the rod is kept at rest in its unloaded straight configuration, which is inclined with respect to an ambient gravitational field. When the system is released, the gravity action provides motion  which is a simple rigid body translation if the rod is highly stiff, say rigid, Fig. \ref{intro} (center). However, when the rod is elastically flexible, transverse (bending) oscillations are displayed and the slip into the sleeve becomes strongly contrasted by the configurational 
force acting on the rod and developed at its insertion point with the sleeve. 
In this latter case, a complex large amplitude dynamic motion develops, during which insertion into the sliding sleeve may alternate with ejection from it, 
thus creating a sort of nonlinear oscillating motion, which eventually terminates with 
the rod either completely ejected from or injected into the sliding sleeve.\footnote{Movies of the experiments can be found in the additional material available at \\ 
\url{http://www.ing.unitn.it/~bigoni/configurationaldynamics.html}
} 

The stroboscopic photos reported in Fig. \ref{intro}
 show the motion of an highly stiff rod (center, realized with a 4 mm thick carbon steel flat strip) and of a flexible rod (right, realized with a 2 mm thick and 25 mm wide carbon-fiber flat strip), both with an initial external length of 580 mm and an attached mass of 1.046 kg. 
 The motion of the rigid and of the flexible  systems ends in both cases with the complete injection of the rod after 0.39 sec and 10.55 sec, respectively (the detailed timing of each snapshot is reported in Table \ref{tabella1}). 
The large difference in the time needed for the two rods to attain their complete injection shows how deeply the dynamics is affected by the elasticity 
of the rod and by the related configurational force action, in the absence of which ejection would never occur! 
Indeed, the configurational force has always the effect of delaying complete injection and even producing ejection.

\begin{table}[H]
	\footnotesize
	\renewcommand{\tablename}{\footnotesize{Tab.}}
	\centering
	\begin{tabu}  to 0.95\linewidth {X[1.02c]| *{4}{X[0.4$c$]}| *{11}{X[0.4$c$]}}
		\toprule
		&  \multicolumn{4}{c|}{Rigid system}      &  \multicolumn{11}{c}{Flexible system} \\
		&  \multicolumn{4}{c|}{(Fig. \ref{intro}, center)}      &  \multicolumn{11}{c}{(Fig. \ref{intro}, right)} \\
		\otoprule
		Snapshot        & 0   & 1   & 2  &  3 & 0 & 1   & 2   & 3   & 4    & 5 &  6  &  7&  8 &  9 & 10 \\	
		Time [s] &\tikz[baseline]{
			\node[anchor=base] (t1)
			{\hspace{-0.15cm}0.00};
		}\;  &  
		\tikz[baseline]{
			\node[anchor=base] (t2)
			{\hspace{-0.15cm}0.13};
		}\; &  
		\tikz[baseline]{
			\node[anchor=base] (t3)
			{\hspace{-0.15cm}0.26};
		}\; &  
		\tikz[baseline]{
			\node[anchor=base] (t4)
			{\hspace{-0.15cm}0.39};
		}\;  & 
		\tikz[baseline]{
			\node[anchor=base] (t5)
			{\hspace{-0.15cm}0.00};
		}\;   &
		\tikz[baseline]{
			\node[anchor=base] (t6)
			{\hspace{-0.15cm}0.44};
		}\;   &  
		\tikz[baseline]{
			\node[anchor=base] (t7)
			{\hspace{-0.15cm}0.66};
		}\;   &  
		\tikz[baseline]{
			\node[anchor=base] (t8)
			{\hspace{-0.15cm}1.11};
		}\;    &  
		\tikz[baseline]{
			\node[anchor=base] (t9)
			{\hspace{-0.15cm}1.44};
		}\;     & 
		\tikz[baseline]{
			\node[anchor=base] (t10)
			{\hspace{-0.15cm}1.95};
		}\;     & 
		\tikz[baseline]{
			\node[anchor=base] (t11)
			{\hspace{-0.15cm}2.14};
		}\;     & 
		\tikz[baseline]{
			\node[anchor=base] (t12)
			{\hspace{-0.15cm}3.63};
		}\;     & 
		\tikz[baseline]{
			\node[anchor=base] (t13)
			{\hspace{-0.15cm}6.64};
		}\;     &
		\tikz[baseline]{
			\node[anchor=base] (t14)
			{\hspace{-0.15cm}8.05};
		}\;     & 
		\tikz[baseline]{
			\node[anchor=base] (t15)
			{\hspace{-0.15cm}10.55};
		}\;   \\
		Motion &	
		\multicolumn{4}{c|}{  }&  \multicolumn{11}{c}{}
	\end{tabu}
	\caption{\footnotesize The timeframe of the snapshots reported in Fig. \ref{intro} (center and right) and type of motion occurring between two successive photos (\lq inject' and \lq eject' stand respectively for injection and ejection). The time 10.55 sec corresponds to the attainment of complete injection of the flexible rod (related snapshot is not reported in Fig. \ref{intro}, for clarity).
	 }
	\label{tabella1}
\end{table}

\begin{tikzpicture}[overlay, auto]  
\node[below right=0.3em and -0.2em of t1.south] (n1) {\footnotesize \color{OliveGreen}inject};
\node[below right=0.3em and -0.2em of t2.south] (n2) {\footnotesize \color{OliveGreen}inject}; 
\node[below right=0.3em and -0.2em of t3.south] (n3) {\footnotesize \color{OliveGreen}inject};      
\end{tikzpicture} 
\begin{tikzpicture}[line width=0.7pt, overlay, auto, OliveGreen] 
\path[->] (t1.south) edge [bend right] (t2.south); 
\path[->] (t2.south) edge [bend right] (t3.south);
\path[->] (t3.south) edge [bend right] (t4.south);
\end{tikzpicture}

\begin{tikzpicture}[overlay, auto]  
\node[below right=0.3em and -0.2em of t5.south] (n4) {\footnotesize \color{OliveGreen}inject};
\node[below right=0.3em and -0.2em of t6.south] (n5) {\footnotesize \color{OliveGreen}inject}; 
\node[below right=0.3em and -0.1em of t7.south] (n6) {\footnotesize \color{Magenta}eject}; 
\node[below right=0.3em and -0.2em of t8.south] (n7) {\footnotesize \color{OliveGreen}inject};    
\node[below right=0.3em and -0.1em of t9.south] (n8) {\footnotesize \color{Magenta}eject};    
\node[below right=0.3em and -0.1em of t10.south] (n9) {\footnotesize \color{Magenta}eject};   
\node[below right=0.3em and -0.1em of t11.south] (n10) {\footnotesize \color{Magenta}eject};    
\node[below right=0.3em and -0.2em of t12.south] (n11) {\footnotesize \color{OliveGreen}inject};    
\node[below right=0.3em and -0.1em of t13.south] (n12) {\footnotesize \color{Magenta}eject};  
\node[below right=0.3em and -0.1em of t14.south] (n13) {\footnotesize \color{OliveGreen}inject};            
\end{tikzpicture} 
\begin{tikzpicture}[line width=0.7pt, overlay, auto, OliveGreen] 
\path[->] (t5.south) edge [bend right] (t6.south); 
\path[->] (t6.south) edge [bend right] (t7.south);
\end{tikzpicture}
\begin{tikzpicture}[line width=0.7pt, overlay, auto, Magenta] 
\path[->] (t7.south) edge [bend right] (t8.south);
\end{tikzpicture}
\begin{tikzpicture}[line width=0.7pt, overlay, auto, OliveGreen] 
\path[->] (t8.south) edge [bend right] (t9.south); 
\end{tikzpicture}
\begin{tikzpicture}[line width=0.7pt, overlay, auto, Magenta] 
\path[->] (t9.south) edge [bend right] (t10.south);
\path[->] (t10.south) edge [bend right] (t11.south);
\path[->] (t11.south) edge [bend right] (t12.south);
\end{tikzpicture}
\begin{tikzpicture}[line width=0.7pt, overlay, auto, OliveGreen] 
\path[->] (t12.south) edge [bend right] (t13.south); 
\end{tikzpicture}
\begin{tikzpicture}[line width=0.7pt, overlay, auto, Magenta] 
\path[->] (t13.south) edge [bend right] (t14.south);
\end{tikzpicture}
\begin{tikzpicture}[line width=0.7pt, overlay, auto, OliveGreen] 
\path[->] (t14.south) edge [bend right] (t15.south); 
\end{tikzpicture}


The length of rod external to the sliding sleeve represents a moving boundary 
(similarly to the boundary conditions encountered in problems involving tensionless surfaces \cite{lenci} and fluid-structure interaction \cite{iutam, elettro1, elettro2}), namely a configurational parameter for the considered structural system. Therefore, a variational technique is used to show that 
a configurational force is generated at the end of the sliding constraint. This force is obtained now within a dynamic context and 
is shown to 
differ from that previously obtained under the quasi-static assumption \cite{bigonieshelby} only for a negligible term
consisting in the ratio between velocity of sliding  and the longitudinal wave speed. 
The dynamic motion shown in Fig. \ref{intro} is analyzed experimentally, with 
a set-up developed at the \lq Instabilities Lab' of the University of Trento, and simulated through a mechanical model in which, essentially, the 
elastic rod is treated as a massless nonlinear spring, obeying the Euler elastica (under the first mode of deformation). 
In agreement with the theoretical predictions, the experiments show that a transition line in a load-inclination plane exists, so that realizations corresponding to 
points within the region above (below) this transition line display final complete  ejection (injection). 

The theoretical and experimental framework introduced in the present article can find application to the design of flexible robot arms with variable 
length or retractable/extensible soft actuators 
\cite{gilbert,kim, rafsa, wang0}. In these devices configurational forces are necessarily generated, providing important effects that cannot be neglected, even in a first approximation design.

\section{The presence of the configurational force disclosed in a dynamic setting}

The presence of a configurational force generated by the sliding sleeve at its exit and proportional to the square of the bending moment at this point was 
demonstrated under the quasi-static assumption \cite{bigonieshelby}. The aim of this section is to theoretically prove that a configurational force  
is generated during a nonlinear dynamic motion and its expression differs from that obtained in the quasi-static setting only in a proportionality coefficient, although this difference  is in practice negligible (because given by the square of the ratio between the sliding velocity and longitudinal wave speed in the bar). The proof is based on the principle of the least action and obtained through a variational technique.

\subsection{Kinematics}

\begin{figure}[!h]
\renewcommand{\figurename}{\footnotesize{Fig.}}
    \begin{center}
    \includegraphics[width=1\textwidth]{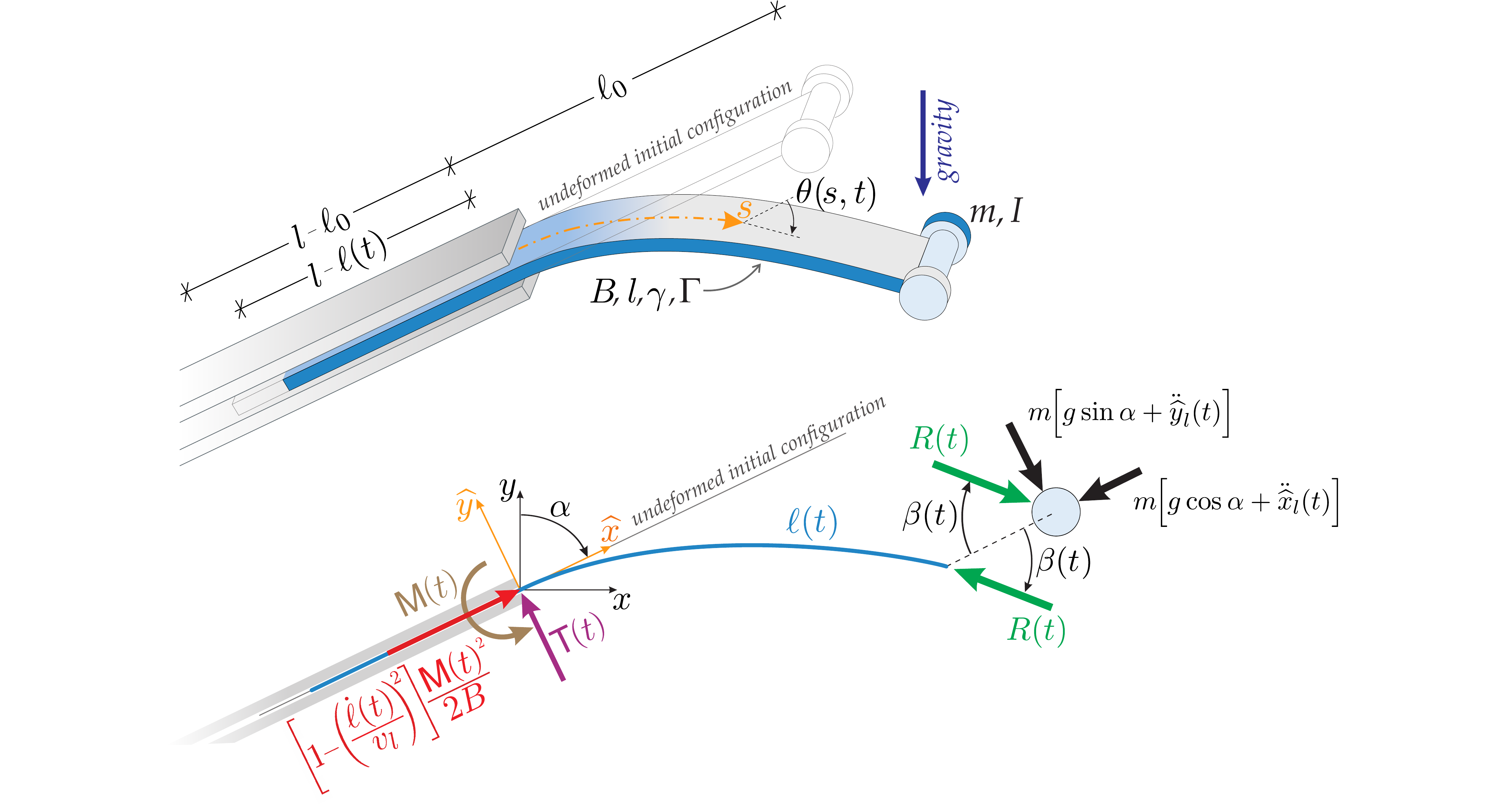}
    \caption{\footnotesize An inextensible linear elastic rod of length $l$, bending stiffness $B$, and linear mass density $\gamma$, with an attached lumped mass $m$ at the end $s=l$ is constrained by a sliding sleeve inclined at an angle $\alpha$ with respect to the $y$ axis. The rotational inertia of the lumped mass is denoted with $I$, while the rod has a uniform rotational inertia density $\Gamma$. The rod is reported in its undeformed initial state (grey) and in a generic deformed configuration (blue) at the time $t$. The latter configuration is defined through the rotation field $\theta(s,t)$ and the configurational parameter $\ell(t)$, defining the amount of rod outside the constraint. Lower part: Free body diagram of the structural system (inertia  forces, distributed along the rod are not reported for simplicity). 
    The three reaction forces developed at the sliding sleeve exit are reported, corresponding to the reaction moment $\mathsf{M}(t)$, the transverse force reaction $\mathsf{T}(t)$, and the configurational force
$[1-(\dot{\ell}(t)/v_l)^2]\mathsf{M}(t)^2/(2B)$, the latter is parallel to the sliding direction (and being $v_l$ the longitudinal wave velocity). }
    \label{System}
    \end{center}
\end{figure}

The kinematics of an inextensible elastic rod of length $l$,  rectilinear in its undeformed configuration, lying within the plane $x-y$ is referred to the (one-dimensional) curvilinear coordinate $s\in[0, l]$ and the time variable $t$. The rod is constrained by a frictionless sliding sleeve, with exit point centered at the coordinates $x=y=0$ and inclined at an angle $\alpha$ with respect to the $y$ axis, Fig. \ref{System}. Considering the primary kinematic field of rotation $\theta(s,t)$,  measuring the clockwise angle with respect to the undeformed rectilinear state, the position fields $x(s,t)$ and $y(s,t)$ can be evaluated from the inextensibility constraint as
\beq
\label{positionfield}
x(s,t) = x(0,t)+\int_{0}^s\sin\left[\theta(s,t)+\alpha\right]\mbox{d}s, ~~~
y(s,t) =y(0,t)+\int_{0}^s  \cos\left[\theta(s,t)+\alpha\right]\mbox{d}s.
\eeq
Because of the presence of the sliding sleeve, in addition to the rotation field $\theta(s,t)$, the structural system is also characterized by the configurational parameter $\ell(t)\in[0,l]$ measuring the length of the rod outside the constraint, defined by the set of points $s\in[l-\ell(t),l]$. Considering the sliding sleeve exit as the origin of the $x-y$ reference system implies the following kinematic constraints for the  position field at the curvilinear coordinate $s=l-\ell(t)$
\beq\label{constraint1}
x\big(s=l-\ell(t),t\big)=y\big(s=l-\ell(t),t\big)=0,
\eeq
while the rotation field of the part of the rod inside of the sliding sleeve remains null, 
\beq\label{constraint2}
\theta(s,t)=0, \qquad s\in[0,l-\ell(t)],
\eeq
so that the coordinates $x(s,t)$ and $y(s,t)$, eqns (\ref{positionfield}), reduce to

\beq
\label{positionreduced}
\resizebox{1\textwidth}{!}{$
x(s,t) =
\left\{
\begin{array}{ll}
\ds -\big(l-\ell(t)-s \big)\, \sin\alpha,\\[4mm]
\ds \int_{l-\ell(t)}^s\sin\left[\theta(s,t)+\alpha\right]\mbox{d}s,
\end{array}
\right.
\qquad
y(s,t) =
\left\{
\begin{array}{ll}
\ds -\big(l-\ell(t)-s \big)\, \cos\alpha,\\[4mm]
\ds \int_{l-\ell(t)}^s  \cos\left[\theta(s,t)+\alpha\right]\mbox{d}s,
\end{array}
\right.
\qquad
\begin{array}{ll}
\ds s\in[0,l-\ell(t)],\\[4mm]
\ds s\in[l-\ell(t),l] .
\end{array}
$}
\eeq

It is also instrumental to consider a further reference system $\widehat{x}-\widehat{y}$, obtained as the counterclowise rotation of the system $x-y$ by the angle $\pi/2-\alpha$, so that the $\hat x$ axis is parallel to the sliding direction. Within this reference system, the rod's kinematics can be described through the position fields $\widehat{x}(s,t)$ and $\widehat{y}(s,t)$ as
\beq
\label{}
\widehat{x}(s,t) =
\left\{
\begin{array}{ll}
\ds -\big(l-\ell(t)-s \big)\, ,\\[4mm]
\ds \int_{l-\ell(t)}^s\cos\theta(s,t)\mbox{d}s
\end{array}
\right.
\qquad
\widehat{y}(s,t) =
\left\{
\begin{array}{ll}
\ds 0,\\[4mm]
\ds -\int_{l-\ell(t)}^s  \sin \theta(s,t) \mbox{d}s,
\end{array}
\right.
\qquad
\begin{array}{ll}
\ds s\in[0,l-\ell(t)],\\[4mm]
\ds s\in[l-\ell(t),l].
\end{array}
\eeq
From the position fields it follows  that the velocity components $\dot{x}(s,t)$ and $\dot{y}(s,t)$ (where a dot represents the time derivative) are given by
\beq
\resizebox{1\textwidth}{!}{$
\dot{x}(s,t) =
\left\{
\begin{array}{ll}
\ds \dot{\ell}(t)\, \sin\alpha,\\[4mm]
\ds \dot{\ell}(t)\, \sin\alpha+\int_{l-\ell(t)}^s \dot{\theta}(s,t)\cos\left[\theta(s,t)+\alpha\right]\mbox{d}s,
\end{array}
\right.
\qquad
\dot{y}(s,t) =
\left\{
\begin{array}{ll}
\ds \dot{\ell}(t)\, \cos\alpha,\\[4mm]
\ds \dot{\ell}(t)\, \cos\alpha-\int_{l-\ell(t)}^s  \dot{\theta}(s,t)\sin\left[\theta(s,t)+\alpha\right]\mbox{d}s,
\end{array}
\right.
\qquad
\begin{array}{ll}
\ds s\in[0,l-\ell(t)],\\[4mm]
\ds s\in[l-\ell(t),l],
\end{array}
$}
\eeq
or, equivalently, in the  $\widehat{x}-\widehat{y}$ reference system  by
\beq\label{velocities}
\resizebox{1\textwidth}{!}{$
\dot{\widehat{x}}(s,t) =
\left\{
\begin{array}{ll}
\ds \dot{\ell}(t)\, ,\\[4mm]
\ds \dot{\ell}(t)+\int_{l-\ell(t)}^s \dot{\theta}(s,t)\sin\theta(s,t)\mbox{d}s
\end{array}
\right.
\qquad
\dot{\widehat{y}}(s,t) =
\left\{
\begin{array}{ll}
\ds 0,\\[4mm]
\ds -\int_{l-\ell(t)}^s  \dot{\theta}(s,t)\cos \theta(s,t) \mbox{d}s,
\end{array}
\right.
\qquad
\begin{array}{ll}
\ds s\in[0,l-\ell(t)],\\[4mm]
\ds s\in[l-\ell(t),l].
\end{array}
$}
\eeq
The time derivative of the null rotation condition (\ref{constraint2}) evaluated at the curvilinear coordinate corresponding to the sliding sleeve exit, $\theta(l-\ell(t),t)=0$, provides the following internal constraint between the time and spatial derivatives (the latter denoted by a prime symbol) of the rotational field at this point through the sliding velocity $\dot{\ell}(t)$ as 
\beq\label{constraintspacetime}
\dot{\theta}\left(l-\ell(t),t\right)=\dot{\ell}\left(t\right)\theta'\left(l-\ell(t),t\right).
\eeq

From eqns (\ref{velocities}) and (\ref{constraintspacetime}) it follows that, while the velocity fields are continuous at the sliding sleeve exit, the velocity of rotation $\dot{\theta}(s,t)$ is spatially discontinuous there because of the discontinuity of the curvature field at the same point, so that the velocity of rotation can be evaluated just inside and outside the sliding sleeve as
\beq
\begin{array}{ll}
\ds \lim_{|\delta|\rightarrow 0}\dot{\theta}\left(l-\ell(t)-|\delta|,t\right)=0,\\
\ds \lim_{|\delta|\rightarrow 0}\dot{\theta}\left(l-\ell(t)+|\delta|,t\right)=\dot{\ell}(t)\lim_{|\delta|\rightarrow 0}\theta'\left(l-\ell(t)+|\delta|,t\right).
\end{array}
\eeq

\subsection{Lagrangian and governing equations}\label{sectionlagrangian}

The Lagrangian functional $\mathcal{L}(t)$ for the considered system is given by
\beq
\begin{array}{ll}
\mathcal{L}(t)=&\ds\mathcal{T}(t)-\mathcal{V}(t)-
\int_{0}^{l} N_x(s,t) \, \left\{x'(s,t)-\sin\left[\theta(s,t)+\alpha\right]\right\} \mbox{d} s\\
&\ds-\int_{0}^{l} N_y(s,t) \, \left\{y'(s,t)-\cos\left[\theta(s,t)+\alpha\right]\right\} \mbox{d}s,
\end{array}
\eeq
where $\mathcal{T}(t)$ is the  kinetic energy, $\mathcal{V}(t)$ is the potential energy, while   $N_x(s)$ and $N_y(s)$ are Lagrangian multipliers (which can be mechanically interpreted as the internal forces along the $x$ and $y$ directions).
Considering that the rod has uniform linear mass density $\gamma$ and rotational inertia density $\Gamma$ (in the Rayleigh sense \cite{graff, piccolroaz}) and has attached a lumped mass $m$ (with rotational inertia $I$) at the coordinate $s=l$, the kinetic energy $\mathcal{T}(t)$  of the system is given by

\beq \mathcal{T}(t)=\frac{m[\dot{x}(l,t)^2+\dot{y}(l,t)^2]}{2}
+\frac{ I \dot{\theta}(l,t)^2}{2}
+\frac{1}{2}\int_{0}^l \gamma
[\dot{x}(s,t)^2+\dot{y}(s,t)^2] \mbox{d}s
+\frac{1}{2}\int_{l-\ell(t)}^l \Gamma
\dot{\theta}(s,t)^2\mbox{d}s. \eeq

The  potential energy $\mathcal{V}(t)$ is  given as the sum of the elastic energy stored inside of the rod and the negative of the work done by the  loads applied to the system.
A quadratic form in the curvature is assumed for the strain energy of the elastic rod,  so that the moment at the coordinate $s$ is given by $M(s,t)=B\theta'(s,t)$, where
$B$ is the (uniform) bending stiffness. Considering a gravitational field characterized by the acceleration $g$ in the direction opposite to the $y$ axis,
the concentrated dead load $P=m g$ is applied at the coordinate $s=l$, while the uniform dead load $\gamma g$ is distributed all along the rod, so that (neglecting an arbitrary constant) the  potential energy $\mathcal{V}(t)$ is given by
\beq
\mathcal{V}(t)= \ds \frac{B}{2}\int_{l-\ell(t)}^l \theta'(s,t)^2 ds
+ P y(l,t) +\int_{0}^{l}\gamma g  y(s,t) \mbox{d}s.
\eeq

The principle of least action can be applied to  the functional $\mathcal{A}$ defined as the integration in time of $\mathcal{L}(t)$
\beq\label{AAAA}
\mathcal{A}=\int_{t_0}^{t^*}  \mathcal{L}(t) \,\,\mbox{d} t,
\eeq
with $t_0$ and $t^*$ being arbitrary initial and final instants of the analyzed time interval. The minimization procedure for the functional $\mathcal{A}$ is expressed by the vanishing of its variation  (see Appendix A for details)
and leads to   the following equations of motion for the part of rod inside the sliding sleeve
\beq\label{gein}
\begin{array}{lll}
 \ds N'_x(s,t)-\gamma \ddot x(s,t)=0,\\[4mm]
\ds N'_y(s,t)-\gamma\left(\ddot y(s,t)+g\right) =0,
\earr
\qquad
s\in[0,l-\ell(t)],
\eeq
 and for the part of rod outside the sliding sleeve
\beq\label{geout}
\begin{array}{lll}
B\theta''(s,t)-\Gamma\ddot\theta(s,t)+N_x(s,t)\cos[\theta(s,t)+\alpha]-N_y(s,t)\sin[\theta(s,t)+\alpha]=0,\\[4mm]
 \ds N'_x(s,t)-\gamma \ddot x(s,t)=0,\\[4mm]
\ds N'_y(s,t)-\gamma\left(\ddot y(s,t)+g\right) =0,
\earr
\qquad
s\in[l-\ell(t),l].
\eeq
Furthermore, as a complement to the differential systems (\ref{gein}) and (\ref{geout}), the minimization procedure also provides the boundary conditions at the two rod's ends
\beq\label{boundcond}
\resizebox{1\textwidth}{!}{$
N_x(0,t)=N_y(0,t)=0,\qquad
N_x(l,t)=-m\ddot x(l,t),\qquad
N_y(l,t)=-m\left(\ddot y(l,t)+g\right),\qquad
M(l,t)=-I\ddot\theta(l,t), 
$}
\eeq
and the interfacial boundary condition at the sliding sleeve exit, $s=l-\ell(t)$,
\beq\label{interfacial}
\salto{0.4}{N_x(l-\ell(t),t)} \sin\alpha+\salto{0.4}{N_y(l-\ell(t),t)}\cos\alpha=\left(1-\frac{\Gamma}{B}\dot{\ell}(t)^2\right)
\frac{\mathsf{M}(t)^2}{2B},
\eeq
where $\mathsf{M}(t)$ is the reaction moment provided by  the sliding sleeve and 
 the symbol
$\salto{0.38}{\cdot}$ denotes the jump in the relevant argument at a specific spatial coordinate, namely
\beq\label{saltodef}
\salto{0.4}{N_j(l-\ell(t),t)}=\lim_{|\delta|\rightarrow 0}\bigg[N_j\big(l-\ell(t)+|\delta|,t\big)-N_j\big(l-\ell(t)-|\delta|,t\big)\bigg],\qquad j=x,y.
\eeq
By considering the moment-curvature linear constitutive relation $M(s,t)=B \theta'(s,t)$ and the condition of null curvature for the part of rod  inside the sliding sleeve,    the reaction moment $\mathsf{M}(t)$  results coincident with the bending moment in the rod evaluated at the moving curvilinear coordinate $s=\lim_{|\delta|\rightarrow 0}(l-\ell(t)+|\delta|)$,
\beq
\mathsf{M}(t)=\lim_{|\delta|\rightarrow 0}B\theta'\big(l-\ell(t)+|\delta|,t\big).
\eeq
The moving  coordinate $s=l-\ell(t)$ is associated with the
cross section at the sliding sleeve exit, so that $\mathsf{M}(t)$ corresponds to the bending moment value at the rod cross section  just outside the constraint.

\subsection{The configurational force in dynamics}

Considering that the internal force components $N_x$ and $N_y$ (along the $x$ and $y$ axes) can be described in terms of the components $N_{\widehat{x}}$ and $N_{\widehat{y}}$
(along the $\widehat{x}$ and $\widehat{y}$ axes)
through the following linear relations
\beq\label{Nhatnohat}
N_x=N_{\widehat{x}}  \sin \alpha - N_{\widehat{y}} \cos \alpha,\qquad
N_y=N_{\widehat{x}}\cos \alpha  + N_{\widehat{y}}\sin \alpha,
\eeq
the jump condition (\ref{interfacial}) at the sliding sleeve can be rewritten as
\beq\label{interfacial2}
\salto{0.4}{N_{\widehat{x}}(l-\ell(t),t)}=
\left(1-\frac{\Gamma}{B}\dot{\ell}(t)^2\right)
\frac{\mathsf{M}(t)^2}{2B},
\eeq
which shows the presence of a non-null jump at the sliding sleeve exit in the internal force component $N_{\widehat{x}}$, representing the internal axial force.
The presence of such a jump in the axial force $N_{\widehat{x}}$ is the result of the action of a configurational force $F_C(t)$ at this point, parallel to the sliding direction ($\widehat{x}$-axis) and equal to
\beq
\label{fconf}
F_C(t)=\left(1-\frac{\Gamma}{B}\dot{\ell}(t)^2\right)
\frac{\mathsf{M}(t)^2}{2B}.
\eeq
It is worth noting that the expression (\ref{fconf}) of the configurational force derived within a dynamic setting differs from that previously obtained under the quasi-static assumption \cite{bigonieshelby} for the proportionality coefficient $1-\Gamma\dot{\ell}(t)^2/B$. Considering that  the density of rotational inertia and the bending stiffness of a rod with homogenous cross section are given by
\beq
\Gamma=\rho J,\qquad
B=E J,
\eeq 
where $E$ is the Young modulus, $J$ is the second moment of the cross section's area,  and $\rho$ is the volumetric density, the configurational force (\ref{fconf}) may be rewritten as
\beq\label{fcdyn}
F_C(t)=\left[1-\left(\frac{\dot{\ell}(t)}{v_l}\right)^2\right]
\frac{\mathsf{M}(t)^2}{2B},
\eeq
where $v_l$ is the longitudinal wave velocity in the rod, $v_l=\sqrt{E/\rho}$. From the mathematical point of view, it follows that:
\begin{itemize}
\item the quasi-static expression $\mathsf{M}(t)^2/(2B)$ represents an upper bound for the configurational force within a dynamic framework,
\beq
F_C(t)\leq\frac{\mathsf{M}(t)^2}{2B},
\eeq
\item excluding supersonic dynamics $(\dot{\ell}(t)>v_l)$, the configurational force always has an outward orientation from the sliding sleeve constraint,
\beq
F_C(t)>0.
\eeq
\end{itemize}
However, in practical applications the sliding velocity of a rod is usually much smaller than its longitudinal wave velocity ($\dot{\ell}(t)\ll v_l$) so that the configurational force is very  well approximated by the expression of its quasi-static counterpart
\beq\label{fcqs}
F_C(t)\approx\frac{\mathsf{M}(t)^2}{2B}.
\eeq

\section{The dynamics of a falling mass attached to an elastic rod sliding through a sleeve}

The dynamic response of the structural system sketched in Figs. \ref{intro} and \ref{System} is now addressed under 
the assumption that the inertia of the system is only provided by the lumped mass $m$, while the remaining inertia 
contributions associated to $\gamma$, $\Gamma$, and $I$ are neglected.  Under this assumption, which is fully 
satisfied in the experimental set-up reported later, the lumped mass  coordinates $x_l(t)=x(l,t)$, $y_l(t)=y(l,t)$ and the configurational parameter $\ell(t)$ represent the  three fundamental kinematic quantities
necessary for describing the evolution in time of the whole mechanical system, because the spatial integration can be independently performed in a closed form \cite{armaninicatapulta}.
Indeed, for rods subject only to concentrated loads, the internal actions $N_x(s,t)$ and $N_y(s,t)$ are piecewise constant in space. In particular, from the integration of  the differential eqns (\ref{gein})$_1$, (\ref{gein})$_2$, (\ref{geout})$_2$, (\ref{geout})$_3$ and 
considering the boundary conditions (\ref{boundcond}), it follows that the part  of the rod inside the sliding sleeve is completely unloaded
\beq
N_{x}(s,t)=N_y(s,t)=0,
\qquad s\in\big[0, l-\ell(t)\big),
\eeq
while the outside part of rod is subject to a constant internal force
\beq
N_{x}(s,t)=\overline{N}_x(t),
\qquad\qquad
N_y(s,t)=\overline{N}_y(t), \qquad\qquad
s\in\big(l-\ell(t),l\big],
\eeq
where
\beq\label{actionnodamp}
\overline{N}_x(t)=-m \ddot{x}_l(t),\qquad
\overline{N}_y(t)=-m \left[ g+\ddot y_l(t)\right].
\eeq
Therefore, the differential eqn (\ref{geout})$_1$ governing the rotation field $\theta(s,t)$ has coefficients varying only in time and can
be rewritten as the \emph{elastica} 
\beq \label{totalona}
\begin{array}{ll}
B \theta''(s,t) + R(t)\sin[\theta(s,t) -\beta(t)] =0,\qquad
s\in[l-\ell(t),l],
\end{array}
\eeq
where $R(t)$ is the resultant force applied at the rod's end, $s=l$, and $\beta(t)$ measures  its clockwise inclination with respect to the $\widehat{x}$ axis,
\beq\label{forzami}
R(t)=\sqrt{\overline{N}_x^2(t)+\overline{N}_y^2(t)},
\qquad
\tan \beta(t)=\frac{\overline{N}_x(t)\cos\alpha-\overline{N}_y(t)\sin\alpha}{\overline{N}_x(t)\sin\alpha+\overline{N}_y(t)\cos\alpha}.
\eeq
Note that, considering relation (\ref{Nhatnohat}), the resultant force $R(t)$  and its inclination $\beta(t)$ can also be expressed in terms of the internal force  components  $N_{\widehat{x}}(t)$ and $N_{\widehat{y}}(t)$ as
\beq\label{forzami2}
R(t)=\sqrt{\overline{N}_{\widehat{x}}^2(t)+\overline{N}_{\widehat{y}}^2(t)},
\qquad
\tan \beta(t)=-\frac{\overline{N}_{\widehat{y}}(t)}{\overline{N}_{\widehat{x}}(t)}.
\eeq

\subsection{Closed-form spatial integration of the elastica}

As previously noticed, because the governing eqn (\ref{totalona}) has coefficients varying only in time, the spatial integration of the elastica can be performed independently of the integration in time, as a function of
the unknown values $R(t)$ and $\beta(t)$. 
 Following \cite{armaninicatapulta}, the spatial integration  of the elastica (\ref{totalona}), 
complemented by the boundary conditions of null rotation at the sliding sleeve
end, $\theta(l-\ell(t),t)=0$, and
of null moment at the rod's end, $\theta'( l,t)=0$, provides the relation between the resultant force $R(t)$, the end rotation $\theta_l(t)$, the resultant inclination $\beta(t)$, and
 the external length $\ell(t)$
\beq
\lb{mu}
R(t) =\frac{ B \big[\mathcal{K}(k(t)) - \mathcal{K}(\sigma_0(t), k(t))\big]^2}{\ell^2(t)},
\eeq
where $\mathcal{K}(k)$ and $\mathcal{K}(\sigma_0,k)$ are, respectively, the complete and incomplete elliptic integrals of the first kind,
\beq
\mathcal{K}(k) = \int\limits_{0}^{\pi/2}\frac{1}{\sqrt{1-k^2\sin^2\phi}}\mbox{d}\phi,
\qquad
\mathcal{K}(\sigma_0,k) = \int\limits_{0}^{\sigma_0}\frac{1}{\sqrt{1-k^2\sin^2\phi}}\mbox{d}\phi,
\eeq
while $k(t)$ and $\sigma_0(t)$ are parameters (varying in time) defined as functions of the rod's end rotation $\theta_l(t)$ and the load inclination $\beta(t)$
as follows
\beq
\lb{Sostituzioni}
k(t) = \sin\left(\dfrac{\theta_l(t)-\beta(t)}{2}\right),
\hspace{1cm}
\sigma_0(t) = -\arcsin \left[\dfrac{1}{k(t)} \sin\left(\dfrac{\beta(t)}{2} \right) \right].
\eeq
Furthermore, the position of the rod's end can be evaluated as
\beq\label{xey}
x_l(t)= \ell(t)\big\{\mathsf{A}(t) \sin[\alpha+\beta(t)] - \mathsf{B}(t) \cos[\alpha+\beta(t)]\big\},\,\,
y_l(t)= \ell(t)\big\{\mathsf{A}(t) \cos[\alpha+\beta(t)] + \mathsf{B}(t) \sin[\alpha+\beta(t)]\big\},
\eeq
or, equivalently, as
\beq\label{xey2}
\widehat{x}_l(t)= \ell(t)\big\{\mathsf{A}(t) \cos \beta(t) +\mathsf{B}(t) \sin\beta(t)\big\},\qquad
\widehat{y}_l(t)= \ell(t)\big\{- \mathsf{A}(t)\sin\beta(t) + \mathsf{B}(t) \cos\beta(t)\big\},
\eeq
where
\beq
\begin{array}{ll}
\mathsf{A}(t)= -1 + \dfrac{2 \bigg\{
\mathcal{E}\big(k(t)\big) - \mathcal{E}(\sigma_0(t),k(t))
\bigg\}}
{\mathcal{K}(k(t)) - \mathcal{K}(\sigma_0(t), k(t))} ,\qquad
\mathsf{B}(t)= - \dfrac{2k(t) \cos \left[\sigma_0(t)\right]}{\mathcal{K}(k(t)) - \mathcal{K}(\sigma_0(t), k(t))} ,
\end{array}
\eeq
and $\mathcal{E}$ is the incomplete elliptic integral of the second kind,
\beq
\mathcal{E}(\sigma_0,k) = \int\limits_{0}^{\sigma_0}\sqrt{1-k^2\sin^2\phi}\,\mbox{d}\phi.
\eeq
From the above equations, the evolution of the elastic system under consideration can be described once the
three fundamental functions $\ell(t)$, $\theta_l(t)$, and $\beta(t)$ are evaluated in time.
The evolution of these three functions is governed by the following nonlinear differential-algebraic equation (DAE) system (where the 
square of the velocity ratio $\dot{\ell}(t)/v_l$ appearing in eqn (\ref{fcdyn}) is assumed negligible)
\begin{equation}
\label{sisnumerico}
\left\{
\begin{array}{lll}
\overline{N}_{\widehat{x}}(t)=-m \left[ g\cos\alpha+\ddot {\widehat{x}}_l(t)\right]-c(t) \dot{\widehat{x}}_l(t),
\\[6mm]
\overline{N}_{\widehat{y}}(t)=-m \left[ g\sin\alpha+\ddot {\widehat{y}}_l(t)\right]-c(t) \dot{\widehat{y}}_l(t),
\\[5mm]
\overline{N}_{\widehat{x}}(t)=-\dfrac{\left[\overline{N}_{\widehat{x}}(t)\widehat{y}_l(t)-\overline{N}_{\widehat{y}}(t)\widehat{x}_l(t)\right]^2}{2B}+\mu \left|\overline{N}_{\widehat{y}}(t)\right| \mbox{sign}\left[\dot{\ell}(t)\right],
\end{array}
\right.
\end{equation}
which is composed by two nonlinear differential equations in the time variable and a nonlinear algebraic equation. The differential equations represent the Newton's second law, eqn (\ref{actionnodamp}),
for the lumped mass, decomposed along the $\widehat{x}$ and $\widehat{y}$ directions. The algebraic equation 
represents the interfacial boundary condition, eqn (\ref{interfacial2}), namely, the axial equilibrium at the sliding sleeve end in the presence of the configurational force. More specifically,
while the former two equations govern the system evolution, the latter provides an implicit relation for the values
assumed by the three functions  $\ell(t)$, $\theta_l(t)$, and $\beta(t)$ at the same instant of time $t$, for example, $\ell(t)=\ell(\theta_l(t),\beta(t))$.
It is  also remarked that the equations of the system (\ref{sisnumerico}) are an enhanced version of those obtained in Sect. \ref{sectionlagrangian}, 
because dissipative effects, essential for comparisons with experiments, are now introduced as follows:
\begin{itemize}
\item the resultant components (\ref{actionnodamp}) acting on the lumped mass are modified into eqns (\ref{sisnumerico})$_1$ and (\ref{sisnumerico})$_2$
to account for viscous dissipation, through
the non-constant parameter $c(t)$ defining a linear damping, related to air drag and to the presence of lubricant in the sliding sleeve. 
Inspired by the definition usually introduced in small amplitude dynamics of rods with fixed length,\footnote{
In the small amplitude dynamics of a lumped mass attached at the free end of a clamped rod (of length $L$ and bending stiffness $B$), the 
following damping coefficient is usually assumed
\begin{equation}
c=2 \zeta\sqrt{\frac{3 m B}{L^3}},
\end{equation}
an expression that can be retrieved from the non-constant parameter $c(t)$, eqn (\ref{noncostantdamping}),  
when the external length $\ell(t)=L$ is assumed constant.}
the non-constant parameter $c(t)$ is assumed, with reference to a constant damping ratio $\zeta$, as
\begin{equation}
\label{noncostantdamping}
c(t)=2 \zeta\sqrt{\frac{3 m B}{\ell(t)^3}};
\end{equation}
\item the axial reaction at the sliding sleeve is modified to account for possible friction forces at the constraint, assumed opposed to the motion and with modulus given by a Coulomb coefficient $\mu$ multiplying the transverse reaction force $\overline{N}_{\widehat{y}}(t)$ at the sliding sleeve.
\end{itemize}

Within a large rotation setting, all the quantities involved in the system (\ref{sisnumerico}) can be expressed as functions
of the three parameters  $\ell(t)$, $\theta_l(t)$, and $\beta(t)$. In particular, through eqn (\ref{xey2}),
the spatial integration of the elastica provides the positions $\widehat{x}_l(t)$ and $\widehat{y}_l(t)$, which derived with respect to time provide the related velocity and acceleration components. Considering now eqns (\ref{forzami2}) and (\ref{mu}), the resultant force components can be written as
\begin{equation}
\label{mu2}
\left\{
\begin{array}{l}
\overline{N}_{\widehat{x}}(t)= - \dfrac{B}{\ell^2(t)}  \big[\mathcal{K}(k(t)) - \mathcal{K}(m(t), k(t))\big]^2 \cos{\beta(t)},
\\[4mm]
\overline{N}_{\widehat{y}}(t)=  \dfrac{B}{\ell^2(t)} \big[\mathcal{K}(k(t)) - \mathcal{K}(m(t), k(t))\big]^2 \sin{\beta(t)}.
\end{array}
\right.
\end{equation}

Due to the strong nonlinearities of both eqns (\ref{sisnumerico}) and (\ref{mu2}), the  integration in time can  only be performed numerically.

\subsection{Initial conditions and numerical time integration}

The motion of the considered structural  system is activated when at the initial time at least one of 
the energies, namely, kinetic $\mathcal{T}(t=0)$ or potential $\mathcal{V}(t=0)$ is non-null, where 
the former energy is related to the mass' velocity and the latter is provided by the sum of both the strain energy stored in the rod 
and the gravitational potential of the lumped mass. To maintain consistency 
with the hypotheses at the basis of the adopted model (which is simplified by assuming a negligibile rod's inertia)\footnote{
The formulated evolutive problem in which the undeformed configuration is assumed at the initial time is consistent with treating the dynamic problem under the assumption of negligible rod's inertia. Differently, initial conditions related to the presence of a non-null curvature at the sliding sleeve exit
may lead to a  jump  in the deformed configuration at the initial time. This would be the result of
a sudden transfer of the configurational force from the sliding sleeve to the point where the external mass is located, $s=l$.} 
and to limit the analysis to initial conditions relatively easy to be experimented, the rod is assumed initially straight 
in its undeformed configuration and at rest. Therefore, the initial kinematics of the lumped mass is given by
\begin{equation}\label{initialconditions}
\widehat{x}_l(0)=\ell_0,\qquad
\widehat{y}_l(0)=\dot{\widehat{x}}_l(0)=\dot{\widehat{y}}_l(0)=0,
\end{equation}
where $\ell_0$ is the length of the rod external to the sliding constraint at the initial time, $\ell_0=\ell(t=0)$. 

Introducing the characteristic time $T$
\begin{equation}
T = \sqrt{\frac{\ell_0}{g}},
\end{equation}
a parametric analysis of the dynamics of the structure in the dimensionless time variable $\tau = t/T$ can be performed 
in terms of the following dimensionless kinematic quantities
\begin{equation}\label{adimensionalizzati}
\lambda(\tau) = \dfrac{\ell(\tau)}{\ell_0}\geq 0,
\qquad
\xi(\tau) = \dfrac{\widehat{x}_l(\tau)}{\ell_0},
\qquad
\eta(\tau) = \dfrac{\widehat{y}_l(\tau)}{\ell_0},
\end{equation}
when the sliding sleeve inclination $\alpha$ and the following load parameter $p$ are varied, 
\begin{equation}\label{adimensionalizzati2}
p = \dfrac{m g\ell_0^2}{B}>0,
\end{equation}
which \lq condenses' both the initial geometrical and loading  conditions.
It follows that the dimensionless version of the initial conditions (\ref{initialconditions}) is given by
\begin{equation}
\label{ciadim}
\xi(0)=1,\qquad
\eta(0)=\overset{*}{\xi}(0)=\overset{*}{\eta}(0)=0.
\end{equation}
where the superimposed \lq *' stands for the derivative with respect to the dimensionless time $\tau$.

The numerical treatment of the equations governing the dynamics of an elastic rod presents certain difficulties connected with the inextensibility constraint \cite{neukirch2012vibrations} and
the strong nonlinearities involved \cite{gazzola}. These difficulties are successfully overcome by  adopting  the numerical technique described in Appendix B and the obtained theoretical predictions are presented in the following subsection.

\subsection{Injection vs. ejection}

A crucial issue in the mechanical behaviour of the structure shown in Figs. \ref{intro} and \ref{System} is to predict if the rod's dynamics will end up with its final complete injection into the sliding sleeve or, oppositely, with its complete ejection from there.

For highly stiff rods ($p\rightarrow 0$), the motion is expected to be a pure rigid translation, described in the absence of dissipative phenomena by
\beq
\lb{pippo}
\frac{\ell^{rigid}(\tau)}{\ell_0}= 1-\frac{\cos\alpha}{2} \tau^2 ,
\eeq
showing that for $\alpha\in (0,\pi/2)$ a rigid rod is always completely injected at the time 
\beq
\tau_{inj}^{rigid} = \sqrt{\frac{2}{\cos \alpha}} ,
\eeq
while for $\alpha\in(\pi/2,\pi)$ a rigid rod is always ejected. By contrast, when  the rod is flexurally deformable, due to the outward direction of the configurational force at the sliding sleeve,  the complete ejection may also be attained for inclinations $\alpha\in (0,\pi/2)$, so that the set of inclinations $\alpha$ corresponding to ejection is enlarged with respect to the rigid case.
 
A theoretical investigation on the motion of the system and the attained final stage is presented below by exploiting the above-developed  model, accounting for dissipative effects. Considering dissipative phenomena in the motion is instrumental for the analysis because the final stage of  complete injection would never be attained for  conservative systems. Such an issue is related to the impossibility of turning the (constant) total potential energy of the system into kinetic energy associated only to a motion parallel to the sliding sleeve direction and at the same time to null elastic energy in the rod.

With reference to  a non-null external length $\ell_0$ and disregarding the rod's mass,
equilibrium configurations have been obtained from a quasi-static analysis \cite{bigonieshelby} as the satisfaction of the following equation, written 
in terms of dimensionless load $p$ and sliding inclination $\alpha$ 
\begin{equation}\label{bosieqn}
p_{eq}(\alpha)= \left[\mathcal{K}\left(\frac{1}{2}\right) + \mathcal{K}\left(\arcsin\left(\frac{1}{\sqrt{2}}\sin\frac{\alpha}{2}\right),\frac{1}{2}\right)\right].
\end{equation}
The condition (\ref{bosieqn}), expressing the geometrical condition of orthogonality between the tangent to the rod's end and the applied load direction, can be approximated
for sliding sleeve inclinations $\alpha\simeq 0$ and $\alpha\simeq \pi/2$ as\footnote{It is worth noting that the (unstable) equilibrium is attained whenever the configurational force balances the component of the load $P$ along the sliding direction, namely $\mathsf{M}^2/(2B)=P \cos\alpha$. Neglecting the nonlinearities provided by large rotations, the moment reaction is approximately expressed by $\mathsf{M}\approx P \ell_0 \sin\alpha$ and therefore the equilibrium is attained when $p\approx 2\cos\alpha/\sin^2\alpha$, an expression that very well approximates eqn (\ref{bosieqn}) for $\alpha\approx \pi/2$. }
\begin{equation}
\label{vac}
p_{eq}(\alpha)\approx \mathcal{K}\left(\frac{1}{2}\right)\left[\mathcal{K}\left(\frac{1}{2}\right)-\sqrt{2}|\alpha|\right]^2+ o(\alpha),\qquad
p_{eq}(\alpha)\approx \pi -2\alpha+ o(\pi -2\alpha)^2.
\end{equation}

The equilibrium configurations solutions of eqn (\ref{bosieqn}) are unstable, so that it is expected that, at least at low dissipation, the dynamic effects originated from any perturbation applied  to the system will push the elastic rod towards
the minimization or the maximization of its external length $\ell(\tau)$ at large times. These two conditions respectively correspond, for a dissipative system,   to a final configuration of complete injection into or complete ejection from the sliding sleeve, 
which can be expressed as
\begin{equation}
\lim_{\tau\rightarrow\infty}\lambda(\tau)=0 \Leftrightarrow
\mbox{complete injection},\qquad \lim_{\tau\rightarrow\infty}\lambda(\tau) \gg 1 
\Leftrightarrow
\mbox{complete ejection}.
\end{equation}
According to the experimental evidence (described in the next section), 
two intervals for $p$ may be identified, which correspond to final states of complete injection and complete ejection,  as
\begin{equation}
\begin{array}{l}\label{inoutinjeje}
p\in\Big(0, p_{tr}(\alpha)\Big) \rightarrow \mbox{complete injection},
\qquad\qquad
p\in\Big(p_{tr}(\alpha),\infty\Big) \rightarrow \mbox{complete ejection},
\end{array}
\end{equation}
where $p_{tr}(\alpha)$ is a \lq transition value' for the dimensionless load at a given sliding sleeve inclination $\alpha$, discriminating between
injection and ejection.
While from a theoretical point of view, an analytical definition of the transition value $p_{tr}(\alpha)$ is not feasible (within a dynamic framework characterized by strong nonlinearities), 
from a practical point of view, the transition can be evaluated by analyzing $\lambda(\tau)$ at large values of the dimensionless time $\tau$. 
This can be done for different values of load $p$ and inclination $\alpha$, as the  result of an integration of the equations of motion  (\ref{sisnumerico}), where dissipative terms are included.
Note that the introduction of the viscous damping ratio $\zeta$, essential to correctly capture the experiments that will be reported in the next section, also 
provides a regularization useful for the stability of the numerical integration. 
The  transition value $p_{tr}(\alpha)$ has been numerically evaluated for a specific inclination angle $\alpha$ through the following iterative procedure based on a bisection method.
At the $j$-th step, the transition value $p_{tr}^{(j)}(\alpha)$ is estimated as
\begin{equation}
p_{tr}^{(j)}(\alpha)=\frac{p_{max, in}^{(j)}(\alpha)+p_{min, out}^{(j)}(\alpha)}{2}, 
\end{equation}
where $p_{max, in}^{(j)}(\alpha)$ and $p_{min, out}^{(j)}(\alpha)$ are respectively  the highest load for the rod injection and the lowest load for ejection, 
both
evaluated at the step $j$.
The iterative procedure is terminated at the step $k$ and the transition value is considered reached, $p_{tr}=p_{tr}^{(k)}$, when its difference with the transition load evaluated in the previous step, $p_{tr}^{(k-1)}$, 
is negligible with respect to a positive threshold $\epsilon_p$
\begin{equation}
\left|\frac{p_{tr}^{(k)}-p_{tr}^{(k-1)}}{p_{tr}^{(k-1)}}\right|< \epsilon_p.
\end{equation}
The numerical integration of the two nonlinear systems has been performed by including the dissipative effects through the parameters $\zeta=0.025$ and $\mu=0.15$. The value of the damping ratio $\zeta$ has been estimated as the average of the equivalent damping ratios providing the logarithmic decrements experimentally measured  during the free oscillations of the system at different, but fixed, length (clamped rod). The value of the friction coefficient $\mu$ has been calibrated through an optimal matching between experimentally measured and theoretical predicted trajectories of the lumped mass (shown in the next section). The numerical integration is accomplished by means of the function $\texttt{NDSolve}$ in \textit{Mathematica} (v. 11) considering the options $\texttt{MaxStepSize}$ $\rightarrow 10^{-3}$,
$\texttt{StartingStepSize}$ $\rightarrow 10^{-8}$, $\texttt{Method}$ $\rightarrow \texttt{IndexReduction}$,\footnote{
The option $\texttt{Method}$ $\rightarrow \texttt{IndexReduction}$ is only needed in the numerical integration during the large rotation regime, namely, in solving  the DAE system (\ref{sisnumerico}). Such an option implies that a differentiation of the algebraic equation is performed during the numerical integration. 
Indeed, being two the indices of the considered DAE system, two differentiations have to be performed on the algebraic equation to successfully perform the integration.} and the small value $\epsilon_{p}=5 \times 10^{-4}$  for the definition of the transition load. 
Such a value for $\epsilon_{p}$ provides  $p_{tr}^{(k)}$ and $p_{tr}^{(k-1)}$, two loads for which the system displays almost the same oscillatory evolution  in the time interval $\tau\in [0,12]$ and within this interval 
does not show tendency towards injection or ejection.

Note that the mechanical problem under consideration is symmetric 
with respect to the inclination $\alpha=0$ (so that the sign of $\alpha$ does not play a role) and the transition load satisfies the property $p_{tr}(\alpha\pm 2m\pi)=p_{tr}(\alpha)$ with $m\in \mathbb{N}_0$. 

The transition load $p_{tr}(\alpha)$ has been 
numerically evaluated for the twentysix inclinations $\alpha$ reported in Table 2 and the corresponding transition curve has been reported in Fig. \ref{ptr}.

\begin{table}[H]
	\footnotesize
	\renewcommand{\tablename}{\footnotesize{Tab.}}
	\centering
	\begin{tabu}  to 1\linewidth {X[2.8c]| *{13}{X[1.6$c$]}}
		\otoprule
		$\alpha\times \dfrac{\pi}{180}$       & 13   & 14   & 15  &  16 & 17 & 18   & 19   &  20   &  21 &  22 &  23 &  24 &  25  \\[1mm]
		\hline
		$p_{tr}$       & 11.870   & 10.715   & 9.700  &  8.814 & 8.030 & 7.324   & 6.753   &  6.378   &  6.039 &  5.675 & 5.331  & 5.016 & 4.731   
	\end{tabu}	
	
	\vspace{3mm}
	
	\begin{tabu}  to 1\linewidth {X[2.8c]| *{13}{X[1.6$c$]}}
		\otoprule
		$\alpha\times \dfrac{\pi}{180}$       &30   & 35   & 40  &  45 & 50 & 55   & 60   &  65   &  70 &  75 &  80 &  85 &  89  \\[1mm]
		\hline
		$p_{tr}$       & 3.599   & 2.813   & 2.239  &  1.806 & 1.471 & 1.208   & 0.995   &  0.817   &  0.672 &  0.544 & 0.429  & 0.337 & 0.201   \\
	\end{tabu}		
	\caption{\footnotesize Transition load $p_{tr}$, 
		discriminating between injection and ejection, numerically evaluated for different sliding sleeve inclinations $\alpha$.}
	\label{tabtabptr}
\end{table}


\begin{figure}[!h]
\renewcommand{\figurename}{\footnotesize{Fig.}}
    \begin{center}
    \includegraphics[width=1\textwidth]{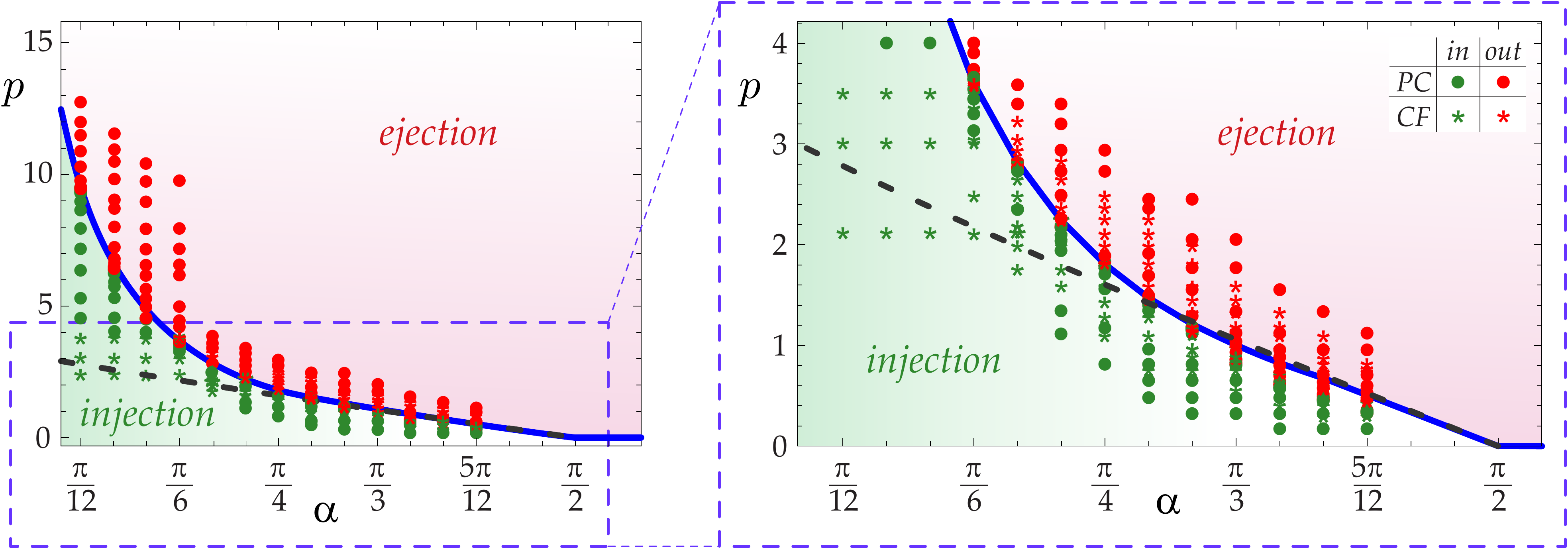}
    \caption{\footnotesize Dimensionless transition load $p_{tr}$ as a function of the  sliding sleeve inclination $\alpha$ (blue curve). The curve separates the two regions for which complete injection ($p < p_{tr}$, green region) and complete ejection ($p \geq p_{tr}$, pink region) respectively occurs as the final state  of the system evolution starting from the undeformed straight configuration. Experimental results (obtained in Sect. \ref{diego}) are also reported for polycarbonate (PC, circle markers) and carbon fibre (CF, star markers) rods, displaying final injection (green markers) and ejection (red markers). The (unstable) equilibrium loading condition $p_{eq}(\alpha)$, eqn (\ref{bosieqn}), is also reported (dashed gray curve).}
    \label{ptr}
    \end{center}
\end{figure}

As described by eqn (\ref{inoutinjeje}), the transition load curve defines two regions within the $p-\alpha$ plane so that the load-inclination pairs lying below/above 
such a curve are related to the final injection/ejection of the elastic rod into/from the sliding sleeve. 
 It can be also observed that:
\begin{itemize}
\item the transition load is a monotonically decreasing function of the sliding sleeve inclination
\beq
\frac{\mbox{d} p_{tr}}{\mbox{d} \alpha}<0;
\eeq
\item injection always occurs for vertical upward sliding sleeve inclinations
\beq
\lim_{\alpha\rightarrow 0} p_{tr}(\alpha)=\infty;
\eeq
\item ejection always occurs for the horizontal sliding sleeve inclination
\beq
\lim_{\alpha\rightarrow \frac{\pi}{2}} p_{tr}(\alpha)=0.
\eeq
\end{itemize}

Note that the loading $p_{eq}(\alpha)$ defined by eqn (\ref{bosieqn}) is also reported dashed in Fig. \ref{ptr}. This load singles out the (unstable) equilibrium configuration of the system for a rod of external length equal to the external length at the initial instant of the motion. The $p_{eq}(\alpha)$  curve is almost 
rectilinear and superimposed to the transition curve for $\alpha\simeq \pi/2$.

Experimental values are also shown in Fig. \ref{ptr} with red and green markers, the former/latter referred to the situation in which complete ejection/injection 
has been observed at the end of the test. Experiments (described in the next Section and performed with rods made up of two different 
materials, polycarbonate and carbon fibre) are in good agreement with the numerical evaluation of the transition curve.

Finally, numerical simulations (not reported for brevity) 
performed with very small values of viscous damping ($\zeta<10^{-2}$), show a highly nonlinear dynamics of the system, numerically difficult to follow. In this case, it is also observed that more than one transition load may be found.

Numerical predictions for the lumped mass trajectory for different sliding inclinations $(\alpha=\{1/12,1/6,1/4,2/3\}\pi)$ are
reported in Fig. \ref{injeject} at increasing load $p=\{0.5,0.999,1.001,1.5\}p_{tr}$.
The simulations show final complete injection for $p<p_{tr}$ and final complete ejection for $p>p_{tr}$. Three deformed configurations of the rod attained during its motion are also reported  
 at different dimensionless time instants
$\tau=\{\tau_1,\tau_2,\tau_4\}$ and  $\tau=\{\tau_1,\tau_2,\tau_3\}$ respectively for $p<p_{tr}$ and for $p>p_{tr}$, with $\tau_1=0.4$,  $\tau_2=2$,$\tau_3=3$, and $\tau_4=4$.

\begin{figure}[!h]
\renewcommand{\figurename}{\footnotesize{Fig.}}
    \begin{center}
    \captionsetup{width=\textwidth}
    \includegraphics[width=1\textwidth]{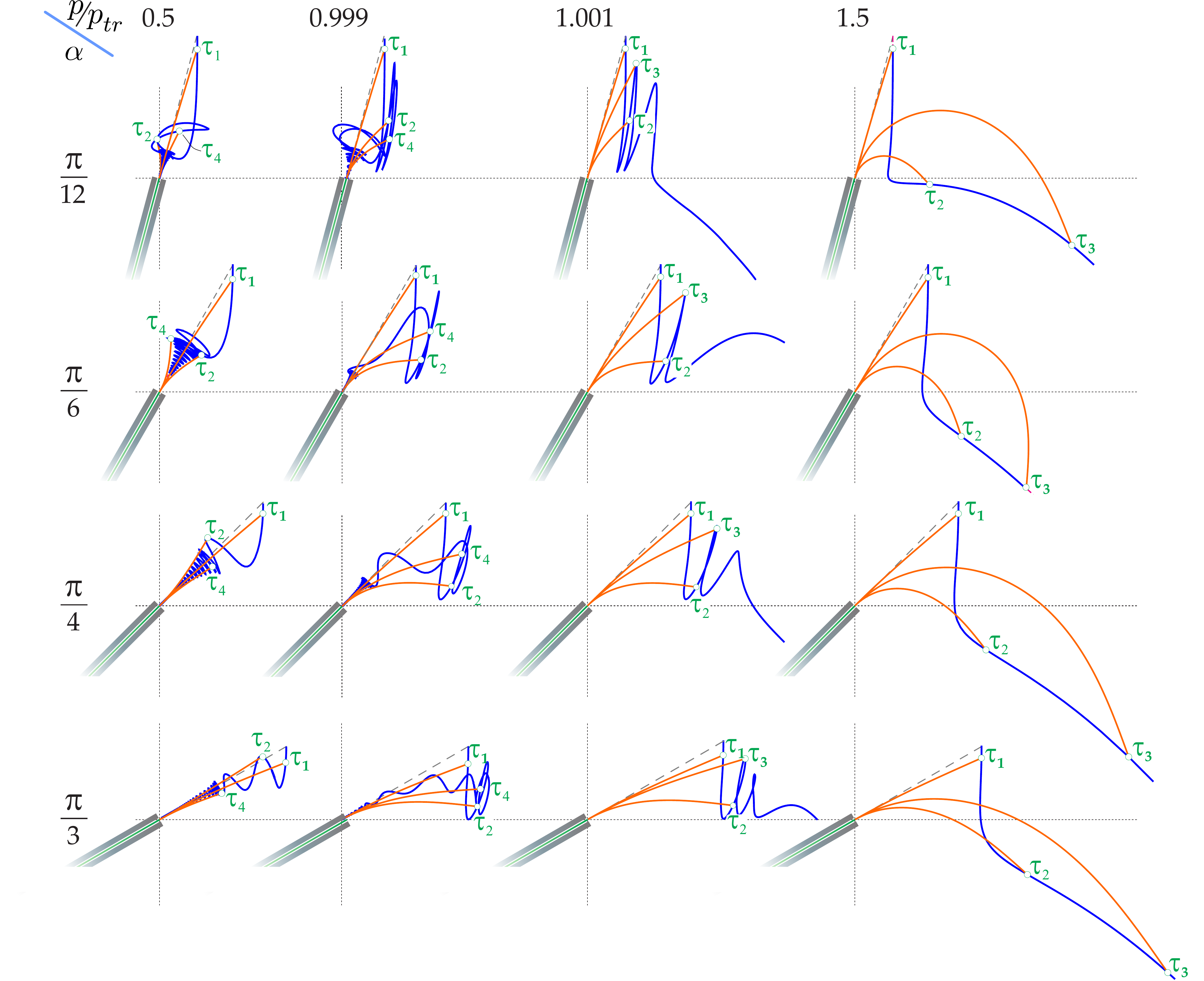}
    \caption{\footnotesize 
		Numerical predictions of the lumped mass trajectory within the dimensionless plane $x/\ell_0-y/\ell_0$ for different sliding sleeve inclinations, $\alpha=\{1/12,1/6,1/4,1/3\}\pi$, and dimensionless loads, $p=\left\{0.5,0.999,1.001,1.5\right\}p_{tr}(\alpha)$. 
Different inclinations (increasing from the upper to the lower part) and different loads (increasing from left to right) are considered. The numerical values of  transition dimensionless loads for the four considered inclinations are reported in Table 2. Deformed configurations attained at specific dimensionless time instants ($\tau_1 = 0.4$, $\tau_2 = 2$, $\tau_3 = 3$, $\tau_4 = 4$) are also reported.}
\label{injeject}
\end{center}
\end{figure}

To provide further insights on the dynamics of the rod, its external length $\lambda(\tau)=\ell(\tau)/\ell_0$ and its phase portrait are reported in Figs. \ref{phase0} and \ref{phase}, for different values of the load $p$ and sliding sleeve inclination $\alpha$. 

Three cases with  $\alpha=\pi/12$ are reported in Fig. \ref{phase0}, one relative to a rigid rod, and the other two corresponding to deformable rods, which exhibit final ejection ($p=1.5 p_{tr}$) and final injection ($p=0.999 p_{tr}$). The evolution of the dimensionless external length  $\ell(\tau)/\ell_0$ is reported as a function of the dimensionless time $\tau$ on the left part of the figure. Here, the rigid system is represented by an half downward parabola (thick red dashed line) with vertex at ($\ell(0)/\ell_0=1,\tau=0$) and intersection with the time axis at $\tau_{inj}^{rigid}=2^{3/4}\sqrt{\sqrt{3}-1}$ (marked as a red diamond marker). The phase portrait diagram is reported on the right part of the figure. Also here, the response of the rigid system is represented by an half-parabola, now leftward, with vertex at ($\ell(0)/\ell_0=1,\dot\ell(0)=0$) and intersection with the velocity axis at $\dot\ell\left(\tau_{inj}^{rigid}\right)=-\sqrt[4]{2+\sqrt{3}}\ell_0/T$.
 The same half parabolas have also been reported  (thin red dashed line) by shifting their vertex to coincide with the four following peaks in the external length $\ell(\tau)$ for the compliant system with $p=0.999 p_{tr}$ (green curves). 

\begin{figure}[!h]
\renewcommand{\figurename}{\footnotesize{Fig.}}
    \begin{center}
    \captionsetup{width=\textwidth}
    \includegraphics[width=1\textwidth]{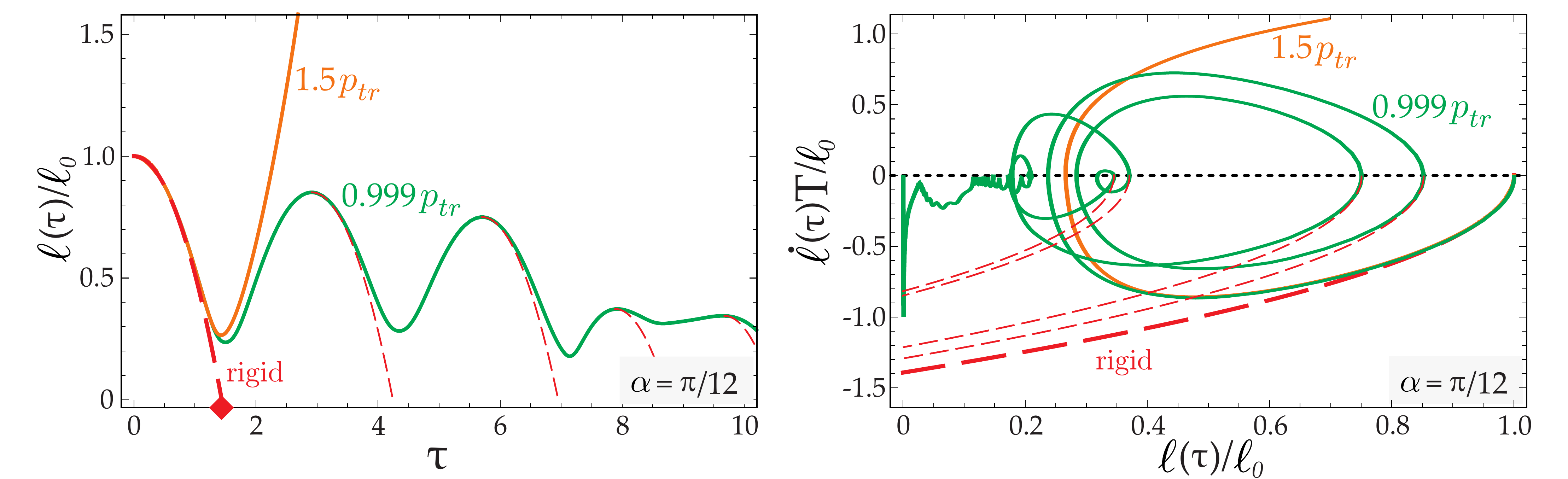}
    \caption{\footnotesize (Left) Evolution of the rod's length external to the sliding sleeve, $\ell$, with the dimensionless time $\tau$ for the rigid system (thick red dashed line) and for two flexible systems, $p=\left\{0.999,1.5\right\}p_{tr}(\alpha)$ (respectively reported as green and orange lines),  and with a sliding sleeve inclination $\alpha=\pi/12$. The  injection time for the rigid system is reported as a red diamond along the dimensionless time axis at $\tau_{inj}^{rigid}=2^{3/4}\sqrt{\sqrt{3}-1}$.
(Right) Phase portrait of $\ell(\tau)$, showing the velocity of the rod's change in length $\dot{\ell}(\tau)$ as a function of $\ell(\tau)$, for the three mechanical systems considered in the left column. Note that 5 half parabolas are reported with a red dashed line, representing 
the rigid-body motion of an infinitely stiff rod with initial conditions corresponding to the vertex of the parabolas.}
    \label{phase0}
    \end{center}
\end{figure}

To better understand the figure,  consider first the response of the compliant system subject to $p=1.5 p_{cr}$ (orange curve). This response is initially very close to that of the rigid system (which ends with the complete injection) and then 
quickly departs from the initial injection stage to ejection 
with an unbounded increase of external length, thus reaching complete ejection.  Considering now the compliant system subject to $p=0.999 p_{tr}$ (green curve), its dynamics is characterized by five oscillations. These oscillations are the repetition of injection stages (closely resembling the rigid-system motion, reported as shifted red dashed parabolas) and ejection stages. 

Similarly to Fig. \ref{phase0}, the dynamic response of the  compliant system is shown in Fig. \ref{phase}  for different
dimensionless loads $p=\left\{0.1,0.5,0.999,1,1.001,1.5,10\right\}p_{tr}(\alpha)$ and  sliding sleeve inclinations, $\alpha=\left\{1/12,1/6,1/4,1/3\right\}\pi$ (increasing from the upper to the lower part).

\begin{figure}[!h]
\renewcommand{\figurename}{\footnotesize{Fig.}}
    \begin{center}
    \captionsetup{width=\textwidth}
    \includegraphics[width=1\textwidth]{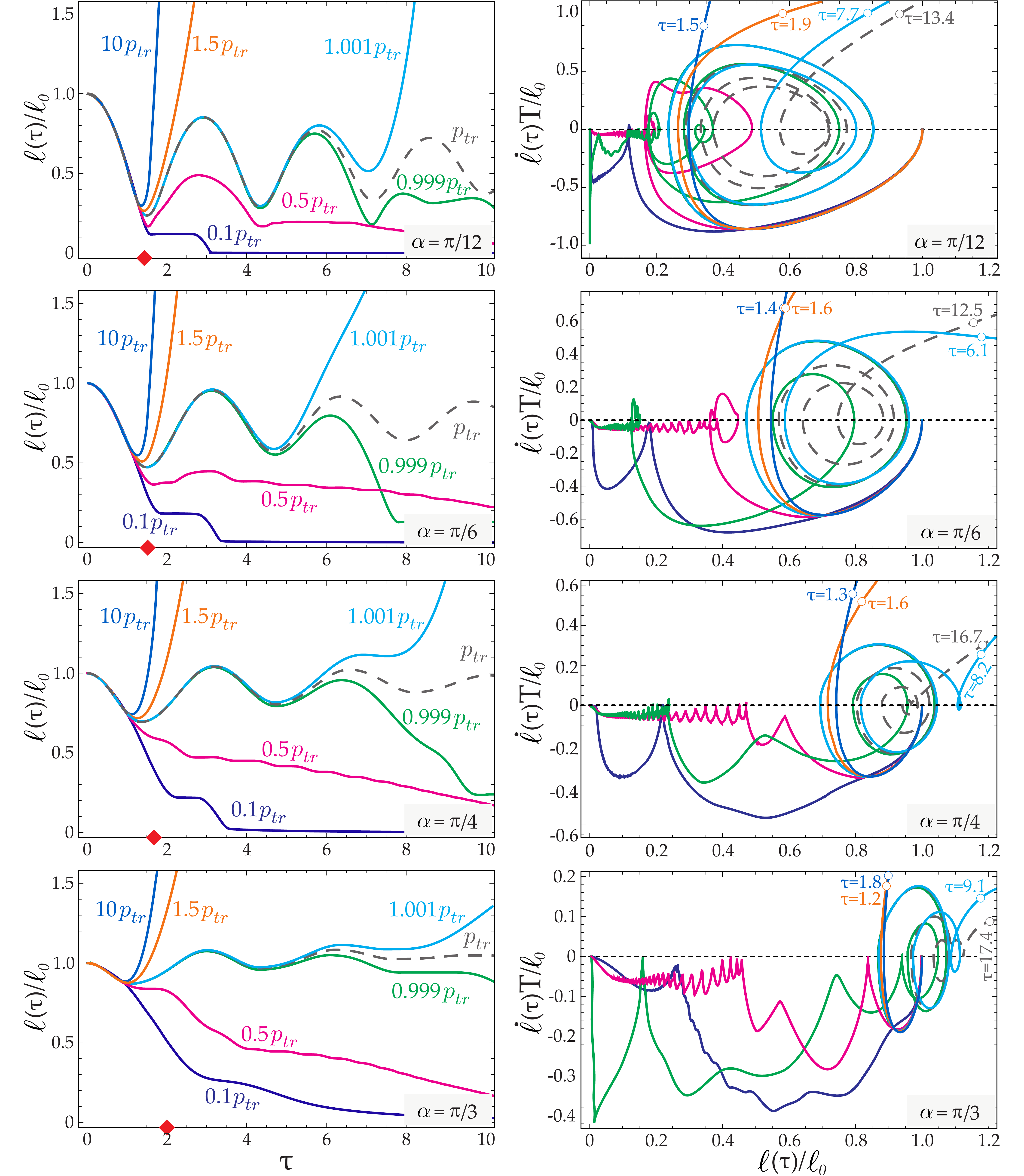}
    \caption{\footnotesize As for Fig. \ref{phase0} but with dimensionless load $p=\left\{0.1,0.5,0.999,1,1.001,1.5,10\right\}p_{tr}(\alpha)$ and at  sliding sleeve inclinations $\alpha=\left\{1/12,1/6,1/4,1/3\right\}\pi$ (increasing from the upper to the lower part). The  injection times for the rigid system at these inclinations are reported as red diamonds along the dimensionless time axis, located at $\tau_{inj}^{rigid}=\{2^{3/4}\sqrt{\sqrt{3}-1},2\times3^{-1/4}, 2^{3/4},2\}$. In the phase portrait the dimensionless time $\tau$ defining states close to the plot-range boundary is also specified for $p=\{1,1.001,1.5,10\}p_{tr}$, 
so that the ejection's acceleration at increasing load can be appreciated. 
		}
    \label{phase}
    \end{center}
\end{figure}

It can be observed from Fig. \ref{phase0}, but even more incisively from Fig. \ref{phase}, that:
\begin{itemize}
\item when $p \simeq p_{tr}$, the system initially displays an oscillatory behaviour with a number of oscillations (before the complete ejection or injection) increasing with the decrease of $|1-p/p_{tr}|$;

\item when $p < p_{tr}$, due to dissipation, the oscillation amplitude decreases in time towards the complete injection of the rod into the sliding constraint;

\item when $p > p_{tr}$, the motion eventually displays an unbounded increase of the external length $\ell(t)$, leading to the complete ejection of the rod from the sliding sleeve.
\end{itemize}

\section{Experiments on the rod's dynamics}
\label{diego}

The experimental setup reported in Fig. \ref{SetupSliding} has been realized and tested (at the Instabilities Lab of the University of Trento) to assess 
the influence of the configurational force generated at the sliding sleeve and to quantify its effect  on the dynamics of the structural system sketched in Figs. \ref{intro} and \ref{System}. 
The experiments, the first of this kind, are complicated by several issues, including the minimization of friction in the sliding sleeve and the realization of 
a triggering mechanism for the instantaneous release of the weight attached to the rod.

\begin{figure}[!h]
\renewcommand{\figurename}{\footnotesize{Fig.}}
    \begin{center}
    \includegraphics[width=1\textwidth]{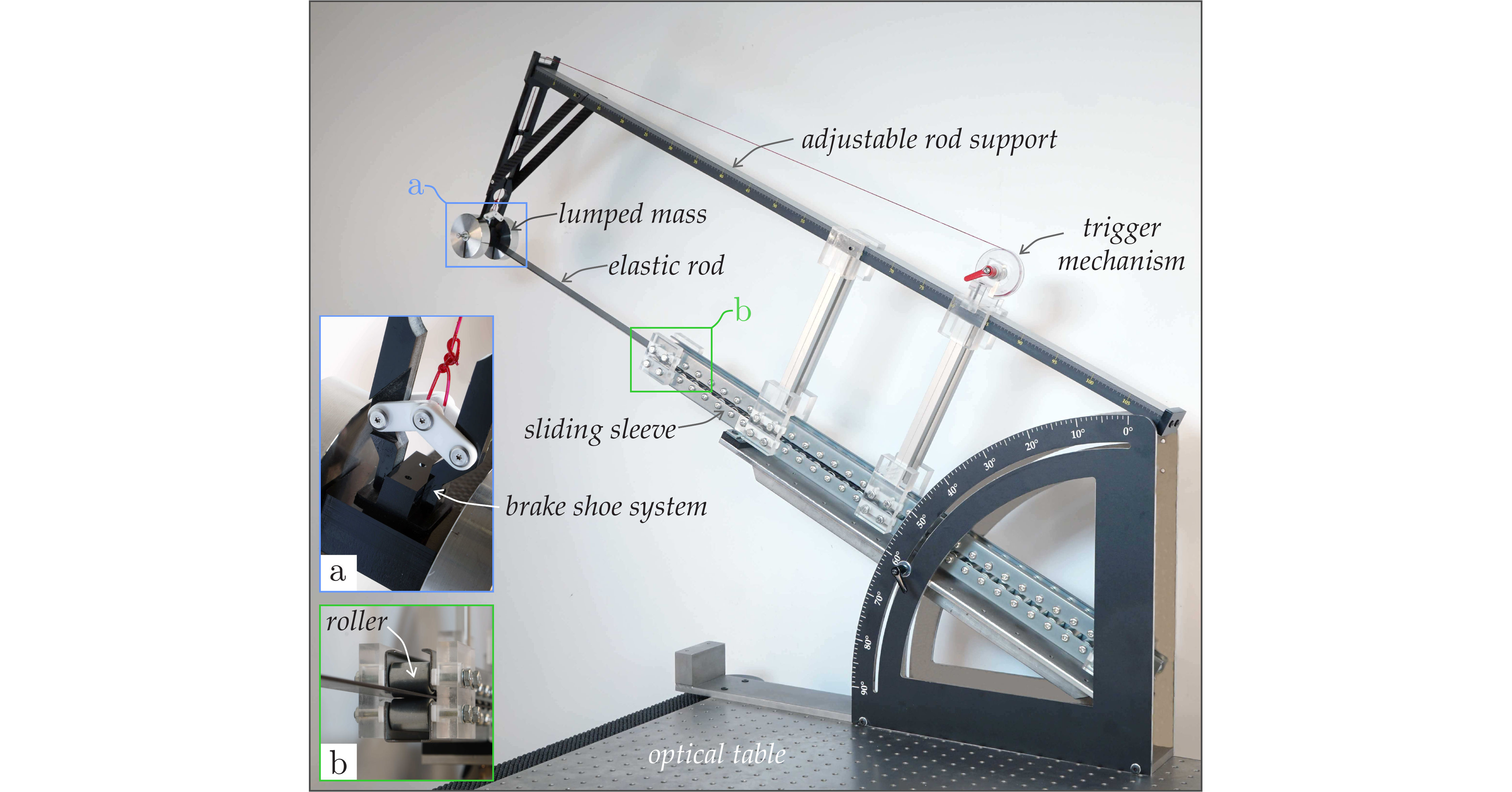}
    \caption{\footnotesize The experimental set-up 
		designed and manufactured to investigate the dynamic response of the structural system shown in Fig. \ref{System}. 
		Insets (a) and (b) report details of the release mechanism
    and of the sliding sleeve, respectively.
    }
    \label{SetupSliding}
    \end{center}
\end{figure}

The sliding sleeve, with an overall length of $825$ mm, was realized with $33$ pairs of rollers  (Fig. \ref{SetupSliding}b).
Each roller is made up of a steel cylinder ($20$ mm diameter and $25$ mm length), containing two roller bearings.
The sliding sleeve was tested under the quasi-static loading and was shown to properly realize the  configurational force predicted by the theoretical model.
Friction was further decreased using a lubricant oil (Ballistol by Klever).

The initial undeformed (straight) condition is realized by holding the rod's end with a brake shoe system (Fig. \ref{SetupSliding}a) placed on a rigid support allowing
to correctly set the initial length of the rod outside the sliding sleeve, $\ell_0$.
The brake shoe system is connected 
through a fishing wire
to a tightener and a trigger, so that it can be easily operated to obtain a sudden release of the weight attached to the rod's end.
The whole apparatus was mounted on a pneumatic optical table (Nexus from ThorLabs), in order to prevent spurious vibrations.

Tests were performed using three different elastic rods, two 
were made up of polycarbonate (PC) strips (Young Modulus $E=2350$ MPa and volumetric mass density $\rho=1180$ kg/m$^3$, so that the longitudinal wave velocity is $v_l\approx 1.4$ km/s), and one was made up of carbon fibre (CF) strips ($E = 80148$ MPa, $\rho=1620$ kg/m$^3$, $v_l\approx 7$ km/s).
The PC rods are both $2.95 \pm 0.05$ mm thick and $25 \pm 0.05$ mm wide, but differ in their length ($550$ mm and $800$ mm).
The CF rod is $2.0 \pm 0.05$ mm thick, $25  \pm 0.05$ mm wide and $800$ mm long.
All the ends of the rods slipping into the sleeve were sharpened using a CNC engraving machine (Roland EGX-600).
A gap 
of $0.5$ mm was always kept between the rod and the rollers along the channel.

High frame-rate movies ($120$ fps) were recorded during each test with a Sony PXW-FS5 video camera, to capture the dynamic motion of the rods. In addition, photos were taken with a Sony Alpha 9 camera.

Several tests were performed on the three rods 
at different inclination angles
$\alpha= -\{1/12, 1/9,$ $ 5/36,$ $1/6, 7/36, 2/9, 1/4, 5/18, 11/36, 1/3, 13/36, 7/18, 5/12, 4/9, 17/36 \}\pi$  for different values 
of the  dimensionless load $p$, through application of different attached masses  $m$ and initial lengths $\ell_0$.\footnote{Experiments 
at inclination angles $\alpha$ smaller than $\pi/12$ were not performed as they were not feasible with the designed experimental set-up, 
which requires long rods or large masses. 
A decrease in the rod stiffness through a reduction in its cross section would cause inelastic deformation in the rod 
during the motion, due to the involved high load levels. It is also worth to mention that the simplified theoretical model, 
based on the negligibility of the rod's mass density, properly predicts 
the experimental observations because all the considered experimental set-ups are characterized by $m>10\gamma l$.}   
 The final stages of complete injection and ejection observed from the experiments for each considered pair $\{p,\alpha\}$ 
are reported  in Fig. \ref{ptr} as green and red symbols, respectively, with circle markers identifying experiments with PC rods and star markers those with CF rods. 

The sliding velocity measured during the experiments
 confirms the smallness of the velocity ratio between rod's sliding and longitudinal wave speed, $|\dot{\ell}(t)/v_l|\lesssim 10^{-3}$, 
and therefore the validity of the approximation 
(\ref{fcqs}) for 
 the configurational force (\ref{fcdyn}).

Experimentally measured injection times $t_{inj}^{exp}$, made dimensionless through division by $t_{inj}^{rigid} = \sqrt{\ell_0/g} \tau_{inj}^{rigid}$, 
are reported in Table \ref{stok} for different sliding sleeve inclinations $\alpha$ and load ratio $p/p_{tr}$. 
\begin{table}[H]
	\footnotesize
	\renewcommand{\tablename}{\footnotesize{Tab.}}
	\centering
	\begin{tabu}  to 0.8\linewidth {X[2.8c]| *{4}{X[1.4$c$]}| *{4}{X[1.4$c$]}}
		\otoprule
			&  \multicolumn{4}{c|}{Polycarbonate (PC)}      &  \multicolumn{4}{c}{Carbon fibre (CF)} \\
		\otoprule
		$\alpha\times \dfrac{\pi}{180}$       & 15   & 30   & 45  &  60 & 15 & 30   & 45   &  60\\[1mm]
		$p/p_{tr}$       & 0.952  & 0.937   & 0.967   &  0.804   & 0.418  & 0.919   & 0.894   &  0.840 \\[1mm] 
		$t_{inj}^{exp}/t_{inj}^{rigid}$       & 8.286  & 23.766   & 20.026   & 31.978     & 16.567  & 23.876 & 27.596   & 38.115 
	\end{tabu}   
	\caption{\footnotesize Experimental measures on polycarbonate and carbon fibre rods of the injection times $t_{inj}^{exp}$, made dimensionless through division by $t_{inj}^{rigid}$, for different sliding sleeve inclinations $\alpha$ and load ratios $p/p_{tr}$. }
	\label{stok}
\end{table}

It can be noticed from the Table that the rod's flexibility, expressed through the load ratio $p/p_{tr}$, increases the injection time substantially, up to an experimentally measured factor of about 38 times the time for injection in the rigid system. The magnification of the injection time provides a further qualitative insight into the 
complexity of the dynamic motion when the flexibility of the system plays a role.

To better appreciate the dynamics of the elastic rod,
stroboscopic photos taken after the mass release are reported in Fig. \ref{Frame_exp}, referred to a sliding sleeve inclination $\alpha=\pi/4$ and loads $p = 0.75 p_{tr}$ (left) and $p = 1.1 p_{tr}$ (right). The mass trajectory is also reported, as measured through an \textit{ad hoc} application developed in Python to process the videos acquired at high frame-rates. More specifically, the code exploits the
OpenCV libraries, a tool-set dedicated to image processing and machine learning applications and, by tracking the trajectory traveled by the lumped mass, provides the position ${x_l,y_l}$ as output for each frame. 
\begin{figure}[!htp]
\renewcommand{\figurename}{\footnotesize{Fig.}}
    \begin{center}
    \includegraphics[width=1\textwidth]{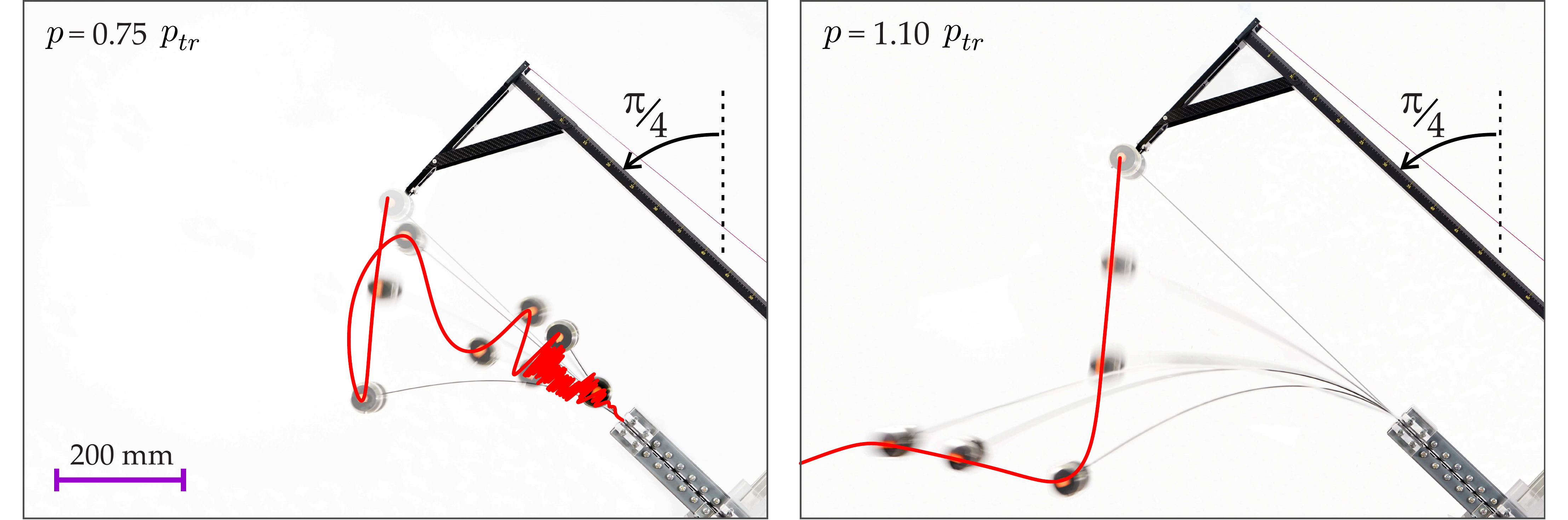}
    \caption{\footnotesize 
The experimental determination of the trajectory of a falling down mass attached to an elastic rod which can slip into a frictionless sliding sleeve. 
Final injection/ejection is observed on the left/right. 
Stroboscopic photos taken during experiments  
(performed with sliding sleeve inclination $\alpha = -\pi/4$ on carbon fibre rods for $p = 0.75 p_{tr}$ on the left and $p = 1.1 p_{tr}$ on the right) are also reported. 
}
    \label{Frame_exp}
    \end{center}
\end{figure}

As a final  comparison between the results of 
mechanical modelling and experiments, 
the trajectory described by the lumped mass and its coordinates evolution in time, $\widehat{x}_l(\tau)/\ell_0$ and $\widehat{y}_l(\tau)/\ell_0$,  after the release from the undeformed configuration are reported in Fig. \ref{TraiettoriaExp}. 
Two settings are considered, $\alpha=\pi/12$ with $p=0.41 p_{tr}$ (upper part) and $\alpha=\pi/4$ with $p=0.90 p_{tr}$ (lower part).
The experimental measures (red curves) are shown to be very well 
predicted by the model (blue curves) and, to better appreciate this, the mass position at the dimensionless times  $\tau_1$, $\tau_2$, $\tau_3$ and $\tau_4$ is highlighted through circles on the  experimental and theoretical trajectory.

\begin{figure}[!htp]
\renewcommand{\figurename}{\footnotesize{Fig.}}
    \begin{center}
    \captionsetup{width=\textwidth}
    \includegraphics[width=1\textwidth]{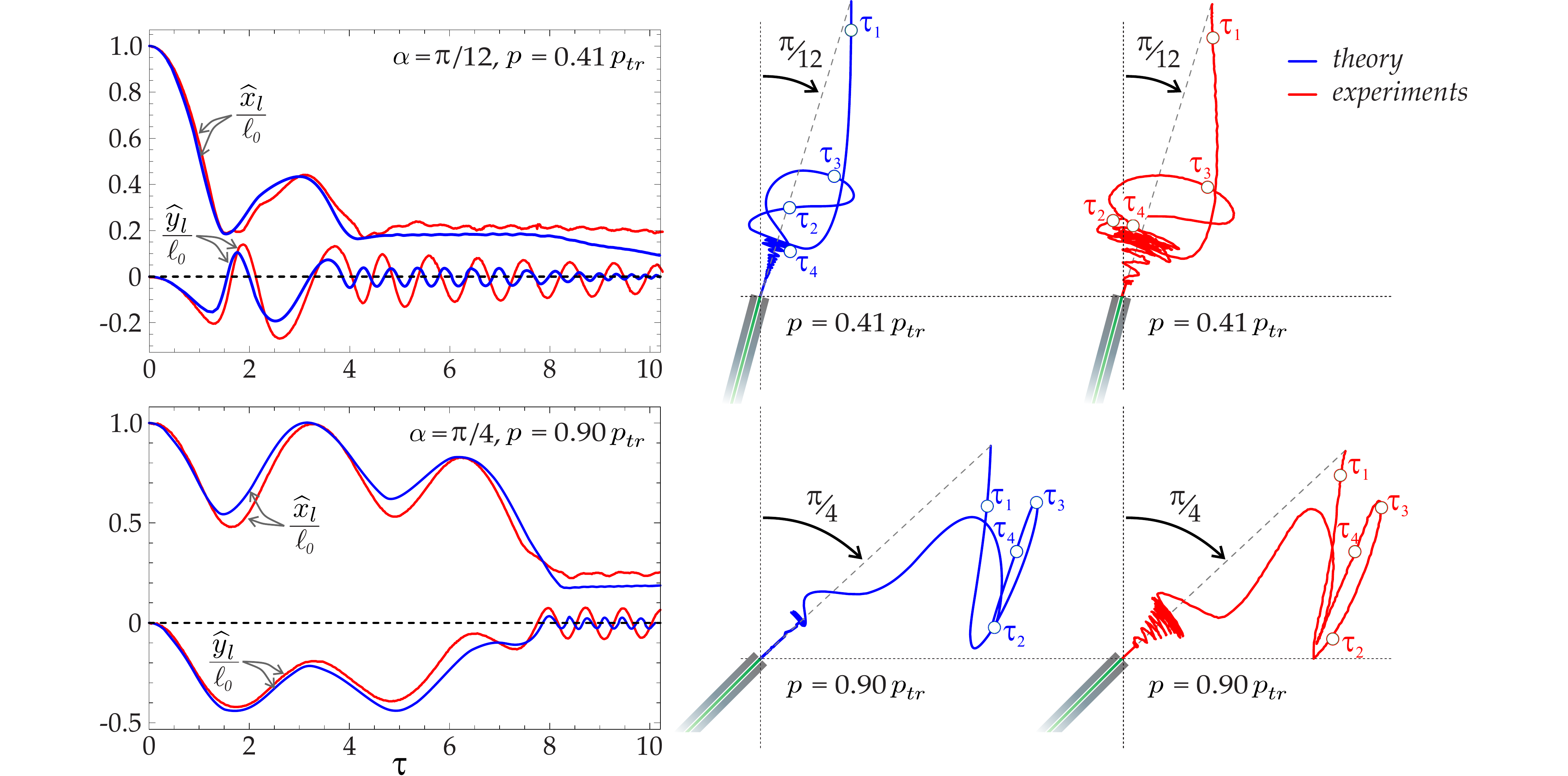}
    \caption{\footnotesize  
The experimental (red curves) and theoretical (blue curves) determination of the trajectory of a falling mass attached to an elastic rod which can slip into a frictionless sliding sleeve. 
		The 		
		mass coordinates $\hat{x}_l$ and $\hat{y}_l$ as functions of the dimensionless time $\tau$ are reported on the left, while the mass trajectory is reported on the right for the following values of parameters: $\alpha = \pi/12$ and  $p = 0.41 p_{tr}$ (upper part), $\alpha =\pi/4 $ and  $p = 0.90 p_{tr}$ (lower part); the rods 
		are made up of carbon fibre strips. 
Final complete injection is attained in both the reported cases. 
		}
    \label{TraiettoriaExp}
    \end{center}
\end{figure}

\section{Conclusions}

The presence of configurational forces during a complex dynamic motion of an elastic rod subject to large deflections 
has been theoretically and experimentally proven, thus extending
previous results restricted to the quasi-static case \cite{bigonieshelby}.
The theoretical proof is based on a variational approach and on the development of a model capable of reproducing the highly nonlinear dynamics of 
an  elastic rod in the presence of: (i.) a sliding sleeve (generating the configurational force), (ii.) a viscous and (iii.) a frictional dissipative term. 
The experimental set-up has permitted the analysis of the rod's dynamic motion and the validation of  the  theoretical model. 

The configurational force, shown to be very well approximated by its counterpart obtained in a quasi-static context, has been demonstrated to represent a decisive ingredient 
in the dynamics of the structure, so that the configurational action generates motions during which, surprisingly, injection alternates with ejection from the sliding sleeve.

The obtained results open new perspectives in the analysis of the dynamic response
of structural systems subject to configurational constraints and may find applications to the mechanical design of innovative flexible devices for soft robotics 
applications.


\paragraph{Acknowledgements.} 
The authors wish to thank Prof. Oliver O'Reilly (University of California, Berkeley) for his 
suggestion of including the rotational inertia of the rod in the article and for disclosing the term 
$\dot{\ell}(t)/v_l$  present in eqn (\ref{fcqs}).
The authors are grateful to Mr. Flavio Vinante (University of Trento) for the invaluable help with the experiments. 
CA, FDC, and DM gratefully acknowledge financial support from the ERC Advanced Grant \lq Instabilities and nonlocal multiscale modelling of materials' (ERC-2013-ADG-340561-INSTABILITIES). 
DB thanks financial support from PRIN 2015LYYXA8-006.

\setcounter{equation}{0}
\renewcommand{\theequation}{{A.}\arabic{equation}}
\section*{Appendix A - Equations of motion from a variational approach}\label{appA}
The equations of motion for the considered system can be obtained through the minimization of the functional $\mathcal{A}$, eqn (\ref{AAAA}). To this purpose the kinematic fields $x(s,t)$, $y(s,t)$, $\theta(s,t)$, and the configurational parameter
$\ell(t)$  are respectively perturbed   by the variation fields  $\epsilon x_{var}(s,t)$, $\epsilon y_{var}(s,t)$, $\epsilon \theta_{var}(s,t)$, and $\epsilon \ell_{var}(t)$,
 where $\epsilon$ is a small quantity, and subject to the following conditions  at the initial and final time,
\beq\label{varcond1}
\begin{array}{lll}
x_{var}(s,t)=y_{var}(s,t)=\theta_{var}(s,t)=\ell_{var}(t)=0,\qquad \mbox{for }\,\,\, t=t_0 \,\,\mbox{and} \,\,t=t^*.
\end{array}
 \eeq

The Taylor series expansion of the kinematical constraint  (\ref{constraint2}) at first-order in the small parameter $\epsilon$  implies the following relation between the variation in the rotation at the sliding sleeve exit $\theta_{var}(l-\ell^-(t),t)$
and the variation in the configurational parameter $\ell_{var}(t)$,
\beq\label{varcond2}
\theta_{var}(l-\ell^-(t),t)=\theta'\big(l-\ell^-(t),t\big)\,\ell_{var}(t),
 \eeq
where the curvilinear coordinate $s=l-\ell^-(t)$ refers to the point just outside the sliding sleeve exit, so that
\begin{equation}
\theta_{var}(l-\ell^-(t),t)=\lim_{|\delta|\rightarrow 0}\theta_{var}\big(l-\ell(t)+|\delta|,t\big),
\qquad
\theta'(l-\ell^-(t),t)=\lim_{|\delta|\rightarrow 0}\theta'\big(l-\ell(t)+|\delta|,t\big).
\end{equation}
Through integration by parts and considering the constraint for the variations expressed by eqn (\ref{varcond2}) and the relation (\ref{constraintspacetime}) between the spatial and the time derivatives of the rotation at the sliding sleeve exit,  the minimization of the functional $\mathcal{A}$, eqn (\ref{AAAA}),  is equivalent to the following expression
\beq\label{varvar}
\begin{array}{lll}
\ds \int_{l-\ell(t)}^{l}\left\{B\theta''(s,t)-\Gamma \ddot\theta(s,t)+N_x(s,t)\cos[\theta(s,t)+\alpha]-N_y(s,t)\sin[\theta(s,t)+\alpha]\right\}\theta_{var}(s,t) \mbox{d}s\\[4mm]
+
\ds  \int_{l-\ell(t)}^{l}\left\{N'_x(s,t)-\gamma \ddot x(s,t)\right\}x_{var}(s,t) \mbox{d}s
+
\int_{l-\ell(t)}^{l}\left\{N'_y(s,t)-\gamma\left(\ddot y(s,t)+g\right) \right\}y_{var}(s,t) \mbox{d}s
\\[4mm]
+
\ds \int_{0}^{l-\ell(t)}\left\{N'_x(s,t)-\gamma \ddot x(s,t)\right\}x_{var}(s,t) \mbox{d}s
+
\int_{0}^{l-\ell(t)}\left\{N'_y(s,t)-\gamma\left(\ddot y(s,t)+g\right) \right\}y_{var}(s,t) \mbox{d}s
\\[4mm]
-
\ds \left\{N_x(l,t)+m \ddot x(l,t)\right\}x_{var}(l,t)
-
\left\{N_y(l,t)+m \left(\ddot y(l,t)+g\right) \right\}y_{var}(l,t)\\[4mm]
+N_x(0,t) x_{var}(0,t)+N_y(0,t)y_{var}(0,t)-\left\{I\ddot\theta(l,t)+B\theta'(l,t)\right\}\theta_{var}(l,t)+\left\{-\salto{0.4}{N_x(l-\ell(t),t)} \sin\alpha\right.
\\[4mm]
 \ds
\left.-\salto{0.4}{N_y(l-\ell(t),t)}\cos\alpha+
\left(B-\Gamma\dot\ell(t)^2\right)\frac{\left[\theta'(l-\ell^-(t),t)\right]^2}{2}
\right\}\ell_{var}(t)
=0.
\earr
\eeq
Imposing the vanishing of this expression for every variation 
field $x_{var}(s,t)$, $y_{var}(s,t)$, $\theta_{var}(s,t)$, and $\ell_{var}(t)$  subject to the conditions (\ref{varcond1}) provides the systems of equations of motion (\ref{gein}) and (\ref{geout}), the boundary conditions (\ref{boundcond}) and the interfacial condition (\ref{interfacial}).
 It is remarked that, in addition to 
the curvature  $\theta'(s,t)$ 
and rotation velocity $\dot{\theta}(s,t)$ fields, the internal forces $N_x(s,t)$ and $N_y(s,t)$ may also have a spatial discontinuity at the sliding sleeve exit $s=l-\ell(t)$, so that
 the symbol
$\salto{0.38}{\cdot}$ has been introduced as the jump in the relevant argument at a specific curvilinear coordinate, as defined in eqn (\ref{saltodef}).
\setcounter{equation}{0}
\renewcommand{\theequation}{{B.}\arabic{equation}}
\section*{Appendix B - The numerical integration strategy}

Because of the inextensibility assumption, a stiffening of the differential system (\ref{sisnumerico}) occurs whenever the rod approaches the undeformed configuration, namely in the small rotation regime.
This stiffening yields a numerical difficulty, which can be overcome by integration under the assumption of small rotation. This approximation is used for the nonlinear system within the time intervals during which 
 the transverse displacement satisfies the following condition
 \begin{equation}
\label{smallregime}
\left|\widehat{y}_l\left(t\right)\right| \, \leq \,\frac{\ell(t)}{200}.
\end{equation}
Within the small rotation regime,  the rod's end position can be computed as
\begin{equation}
\label{approxxx1}
\widehat{x}_l(t) = \ell(t),
\qquad
\widehat{y}_l(t) =\dfrac{\overline{N}_{\widehat{y}}(t)\ell^3(t)}{3 B},
\end{equation}
and the following  moment inequality holds
\begin{equation}
\label{approxxx2}
\left|\overline{N}_{\widehat{x}}(t)\widehat{y}_l(t)\right|\ll
\left|\overline{N}_{\widehat{y}}(t)\widehat{x}_l(t)\right|,
\end{equation}
so that the nonlinear DAE system (\ref{sisnumerico})
can be reduced to the following system of two nonlinear differential equations
\begin{equation}
\label{sisnumerico2}
\left\{
\begin{array}{lll}
\dfrac{9B}{2} \dfrac{ {\widehat{y}}^2_l(t)}{{\widehat{x}}^4_l(t)}
-3B \mu\left|\dfrac{\widehat{y}_l(t)}{\widehat{x}^3_l(t)}\right|
\mbox{sign}\left[\dot{\widehat{x}}_l(t)\right]
=m \left[ g\cos\alpha+\ddot {\widehat{x}}_l(t)\right]+c \dot{\widehat{x}}_l(t),
\\[6mm]
3B \dfrac{\widehat{y}_l(t)}{\widehat{x}^3_l(t)}=-m \left[ g\sin\alpha+\ddot {\widehat{y}}_l(t)\right]-c \dot{\widehat{y}}_l(t).
\end{array}
\right.
\end{equation}
It is remarked that due to the small rotation assumption, the nonlinear system (\ref{sisnumerico2}) is solved in terms of two
independent quantities $\widehat{x}_l(t)=\ell(t)$ and $\widehat{y}_l(t)$ only, a procedure much easier than solving the system (\ref{sisnumerico}), where $\ell(t)$ is not constrained to be equal to $\widehat{x}_l(t)$.
For this reason, the performed numerical integration introduces a discontinuity in $\ell\left(\tilde{t}_i\right)$ and $\dot\ell\left(\tilde{t}_i\right)$ at all the
\lq passage'  times $\tilde{t}_i$ for which
 \begin{equation}
\label{costanza}
\left|\widehat{y}_l\left(\tilde{t}_i\right)\right|\, \simeq \,\frac{\ell\left(\tilde{t}_i\right)}{200},\qquad i\in\mathbb{N}, 
\end{equation}
corresponding to the passage between the large and small rotation regimes and implying the passage from the integration of the nonlinear  system (\ref{sisnumerico}) to its approximated version (\ref{sisnumerico2}) and viceversa.

In particular, the integration starts at $t=0$ within the small rotation regime, so that the external length is constrained to be equal to the axial position, $\ell(t)=\widehat{x}_l(t)$, and
therefore its velocity is given by $\dot\ell(t)=\dot{\widehat{x}}_l(t)$, two conditions which do not need to be satisfied when the small rotation condition (\ref{smallregime}) is no longer satisfied.

The value of $\ell\left(\tilde{t}_1\right)$ and the corresponding velocity $\dot\ell\left(\tilde{t}_1\right)$  at the first passage time $t=\tilde{t}_1$ can be obtained
from eqn (\ref{sisnumerico})$_3$ and its time derivative. It follows that jumps in the external length $\ell(t)$ and its velocity $\dot\ell(t)$
are originated at the first passage time $\tilde{t}_1$ 
and similarly at all the passage times  $\tilde{t}_i$ ($i\in\mathbb{N}$), when the numerical integration switches from the
nonlinear system, eq (\ref{sisnumerico}), to its approximation, eq (\ref{sisnumerico2}).
Nevertheless, it is shown below that the jumps in the mentioned quantities are always found to be negligible because the passage from one of the two solving systems to the other occurs when the condition (\ref{costanza}) is verified.

The jumps 
generated by the adopted numerical strategy are monitored to be negligible by checking the following inequalities for all the passage times $\tilde{\tau}_i$
\beq
\frac{|\ell(\tilde{\tau}_i^+)-\ell(\tilde{\tau}_i^-)|}{\ell(\tilde{\tau}_i^-)}<\epsilon_\ell,
\qquad
\frac{\left|\overset{*}{\ell}(\tilde{\tau}_i^+)-\overset{*}{\ell}(\tilde{\tau}_i^-)\right|}{\overset{*}{\ell}(\tilde{\tau}_i^-)}<\epsilon_{\overset{*}{\ell}},\qquad i\in \mathbb{N},
\eeq
being $\epsilon_\ell$ and $\epsilon_{\overset{*}{\ell}}$ positive but small values.
Moreover, the consistency of the numerical integration is assessed through a comparison between  the total energy decrease $\Delta \Pi(\tau)$ of the system
\begin{equation}
\Delta \Pi(\tau)=\mathcal{T}(\tau)+\mathcal{V}(\tau)-\mathcal{V}(0)<0,
\end{equation}
and the work dissipated $\mathcal{W}_d(\tau)$ through viscosity and friction, 
\begin{equation}
\resizebox{1\textwidth}{!}{$
\mathcal{W}_d(\tau) =\dfrac{B}{\ell_0} \ds \int_{0}^\tau 
\left\{
2 \zeta  \sqrt{\frac{3 p}{\lambda(\tau)^3}}
\left\{
\left[\overset{*}{\xi}(\tau)\right]^2+
\left[\overset{*}{\eta}(\tau)\right]^2\right\}
+
\mu
\left|\overset{*}{\lambda}(\tau)
\left[
2 \zeta  \sqrt{\frac{3 p}{\lambda(\tau)^3}}\overset{*}{\eta}(\tau)
+p\left(\overset{**}{\eta}(\tau)+\sin\alpha\right)
\right]
\right|\right\}\mbox{d}\tau,
$}
\end{equation}
so that the following condition for the normalized modulus of their difference is verified at every integration time $\tau$ 
\begin{equation}
\left\vert \dfrac{\Delta \Pi(\tau)- \mathcal{W}_d(\tau) }{\Delta \Pi(\tau)} \right\vert <\epsilon_{\mathcal{W}},
\end{equation}
being $\epsilon_{\mathcal{W}}$ a positive, but small, value. More specifically, the reported simulations 
are obtained by using the following small positive values, which define the precision in the numerical integration scheme,
\beq
\epsilon_\ell= 5 \times 10^{-5},\qquad
\epsilon_{\overset{*}{\ell}}= 10^{-2},\qquad
\epsilon_{\mathcal{W}}=2 \times 10^{-4}. 
\eeq


\begin{thebibliography}{99}

\setlength{\itemsep}{-1.0mm}

\bibitem{ada}
Amendola, A., Krushynska, A. Daraio, C. Pugno, N.M., Fraternali, F. (2018) Tuning frequency band gaps of tensegrity mass-spring chains with local and global prestress. \emph{Int. J. Sol. Struct.},  155, 47--56.

\bibitem{armaninicatapulta}
Armanini, C., Dal Corso, F. , Misseroni, D., Bigoni, D. (2017)
From the elastica compass to the elastica catapult: an essay on the mechanics of soft robotic arm.
\emph{Proc. R. Soc. A}, 473, 20160870.

\bibitem{ballarini}
Ballarini, R., Royer-Carfagni, G. (2016) A Newtonian interpretation of configurational forces on dislocations and cracks. \emph{J. Mech. Phys. Sol.}, 95, 602-620 

\bibitem{bigonitorsionalgun}
Bigoni, D., Dal Corso, F., Misseroni, D., Bosi, F. (2014)
Torsional locomotion
\emph{Proc. R. Soc. A}, 470.2171, 20140599.

\bibitem{bigoniblade}
Bigoni, D., Bosi, F., Dal Corso, F., Misseroni, D. (2014)
Instability of a penetrating blade
\emph{J. Mech. Phys. Solids}, 64, 411--425.

\bibitem{bigonieshelby}
Bigoni, D., Dal Corso, F., Bosi, F., Misseroni, D. (2015)
Eshelby-like forces acting on elastic structures: theoretical and experimental proof.
\emph{Mech. Mater.}, 80, 368--374.

\bibitem{bosiarmscale}
Bosi, F., Misseroni, D., Dal Corso, F., Bigoni, D. (2014)
An elastica arm scale.
\emph{Proc. R. Soc. A}, 470, 20160870.

\bibitem{bosidripping}
Bosi, F., Misseroni, D., Dal Corso, F., Bigoni, D. (2015).
Self-encapsulation, or the \lq dripping' of an elastic rod.
\emph{Proc. R. Soc. A}, 471, 20150195.



\bibitem{carta}
Carta, G.,  Jones, I.S., Movchan, N.V., Movchan, A.B.,  Nieves, M.J.  (2017) Gyro-elastic beams for the vibration reduction of long flexural systems. \emph{Proc. R. Soc. A}, 473, 20170136.


\bibitem{cazzolli}
Cazzolli, A., Dal Corso, F. (2019)
Snapping of elastic strips with controlled ends.
\emph{Int. J. Sol. Struct.},  162, 285-303.


\bibitem{lenci} Demeio, L., Lancioni, G., Lenci, S. (2011) Nonlinear resonances in infinitely long 1D continua on a tensionless substrate. \emph{Nonlinear Dynamics}, 66, 271--284. 


\bibitem{dalcorsosnake}
Dal Corso, F., Misseroni, D., Pugno, N.M., Movchan, A.B., Movchan, N.V., Bigoni, D. (2017) 
Serpentine locomotion through elastic energy release.
\emph{J. R. Soc. Interface}, 14, 20170055.

\bibitem{elettro1}
Elettro, H., Vollrath, F., Antkowiak, A., Neukirch, S. (2017) Drop-on-coilable-fibre systems exhibit negative stiffness events and transitions in coiling morphology.
\emph{Soft Matter}, 13, 5509-5517.

\bibitem{elettro2} Elettro, H., Grandgeorge, P., Neukirch, S. (2017) Elastocapillary coiling of an elastic rod inside a drop. \emph{J. Elasticity}, 127, 235-247. 


\bibitem{eshelby1} Eshelby, J.D. (1951) The force on an elastic singularity. \emph{Phil. Trans. R. Soc. A}, 244-877, 87-112.

\bibitem{eshelby2} Eshelby, J.D. (1956) The continuum theory of lattice defects. \emph{Solid State Phys.}, 3.C, 79-144.

\bibitem{eshelby3}
Eshelby, J.D. (1970)
Energy relations and the energy-momentum tensor in continuum mechanics.
in \emph{Inelastic Behaviour of Solids},  (eds. M. Kanninien, W. Adler, A. Rosenfield, and R. Jaffee),  77--115, McGraw-Hill, New York.

\bibitem{eshelby4}
Eshelby, J.D. (1975)
The elastic energy-momentum tensor.
\emph{Journal of Elasticity}, 5.3-4,  321--335.


    
\bibitem{garau}
Garau, M, Carta, G, Nieves, M.J, Jones, I.S, Movchan, N.V., Movchan, A.B. (2018) Interfacial waveforms in chiral lattices with gyroscopic spinners. \emph{Proc. R. Soc. A}, 474, 20180132.

\bibitem{gazzola}
Gazzola, M., Dudte, L.H., McCormick, A.G., Mahadevan, L. (2018) Forward and inverse problems in the mechanics of soft filaments. \emph{R. Soc. Open Sci.}, 5, 171628.

\bibitem{gilbert}
Gilbert, H.B., Rucker, D.C., Webster, R.J. (2016) Concentric tube robots: The state of the art and future directions. \emph{Springer Tracts in Advanced Robotics}, 114, 253--269.

\bibitem{gomez}
Gomez, M., Moulton, D.E., Vella, D.
(2019) Dynamics of viscoelastic snap-through. 
\emph{J. Mech. Phys. Sol.}, 124, 781--81.



\bibitem{graff}
Graff, K.F. (1991) Wave Motion in Elastic Solids. Dover, New York.



\bibitem{gravagne}
Gravagne, I.A., Rahn,  C.D., Walker, I.D. (2003) Large deflection dynamics and control for planar continuum robots. \emph{IEEE/ASME Trans. Mechatronics}, 8 (2), 299-307.

\bibitem{hanna}
Hanna, J.A., Singh, H., Virga, E.G. (2018) Partial Constraint Singularities in Elastic Rods. \emph{J. Elas.}, 133, 105-118.

\bibitem{iutam}  Gutschmidt, S., Hewett, J.N., Sellier, M. (2019) IUTAM Symposium on Recent Advances in Moving Boundary Problems in Mechanics - Proceedings of the IUTAM Symposium on Moving Boundary Problems, Christchurch, New Zealand, February 12-15, 2018. Springer. 


\bibitem{kim}
Kim, S., Laschi, C., Trimmer, B. (2013) Softrobotics: a bioinspired evolution in robotics. \emph{Trends. Biotechnol.}, 31, 287–294.


\bibitem{koch}
Kochmann, D., Bertoldi, K. (2017) Exploiting Microstructural Instabilities in Solids and Structures: From Metamaterials to Structural Transitions. \emph{Appl. Mechanics Rev.}, 69, 050801.

\bibitem{liakou1} Liakou, A. (2018) Constrained buckling of spatial elastica: Application of optimal control method. \emph{J. App. Mech. ASME}, 85, 081005.

\bibitem{liakou2} Liakou, A. (2018) Application of optimal control method in buckling analysis of constrained elastica problems. \emph{ Int. J. Sol. Struct.}, 141-142, 158-172.

\bibitem{liakou3} Liakou, A., Detournay, E. (2018) Constrained buckling of variable length elastica: Solution by geometrical segmentation. \emph{Int. J. Non-Linear Mech.}, 99, 204-217. 

\bibitem{nadkarni}
Nadkarni, N., Arrieta, A.F., Chong, C., Kochmann, D.M., Daraio, C. (2016) 
Unidirectional transition waves in bistable lattices. \emph{Phys. Rev. Lett.}, 116, 244501.

\bibitem{neukirch2012vibrations}
Neukirch, S., Frelat, J., Goriely, A., Maurini, C. (2012)
Vibrations of post-buckled rods: the singular inextensible limit. 
\emph{J. Sound Vib.}, 331 (3), 704--720.

\bibitem{nievesJMPS} 
Nieves, M.J., Carta, G., Jones, I.S., Movchan A.B.,  Movchan, N.V. (2018) Vibrations and elastic waves in chiral multi-structures. \emph{J. Mech. Phys. Solids}, 121, 387--408.


\bibitem{oreilly1} 
O'Reilly, O.M. (2015) Some perspectives on Eshelby-like forces in the elastica arm scale. \emph{Proc. R. Soc. A}, 471, 20140785. 

\bibitem{oreilly2} 
O'Reilly, O. (2017) Modeling Nonlinear Problems in the Mechanics of Strings and Rods: The Role of the Balance Laws. Springer. 


\bibitem{pandey}
Pandey, A., Moulton, D.E., Vella, D., Holmes, D.P. (2014)
Dynamics of snapping beams and jumping poppers. \emph{Eur. Phys. Lett.} 105 (2): 24001.

\bibitem{piccolroaz} 
Piccolroaz, A., Movchan, A.B. (2014)
Dispersion and localization in structured Rayleigh beams.
\emph{Int. J. Sol. Struct.}, 51, 4452--4461.



    
\bibitem{rafsa}
Rafsanjani, A., Zhang, Y., Liu, B., Rubinstein, S.M., Bertoldi, K. (2018) Kirigami skins make a simple soft actuator crawl. \emph{Science Robotics} 3 (15), eaar7555.

\bibitem{renda}
Renda, F., Giorelli, M., Calisti, M., Cianchetti, M., Laschi, C. (2014) Dynamic Model of a Multibending Soft Robot Arm Driven by Cables. \emph{IEEE Trans. Robotics}, 30 (5), 1109--1122.

\bibitem{singh} Singh, H., Hanna, J.A. (2019) On the Planar Elastica, Stress, and Material Stress. \emph{J. Elas}, 136 (1), 87--101.

\bibitem{sugino}
Sugino, C., Ruzzene, M., Erturk, A. (2018) Merging mechanical and electromechanical bandgaps in locally resonant metamaterials and metastructures. \emph{J. Mech. Phys. Sol.}, 116, 323--333. 


\bibitem{tsuda}
Tsuda, T., Mochiyama, H. Fujimoto, H.  (2012)
Quick stair-climbing using snap-through buckling of closed elastica. \emph{Int. Symp. Micro-NanoMechatr. Human Sci., MHS 2012}, 368--373.

\bibitem{wang0}
Wang, Z., Polygerinos, P., Overvelde, J.T.B., Galloway, K.C., Bertoldi, K., Walsh, C.J. (2017) Interaction Forces of Soft Fiber Reinforced Bending Actuators. \emph{IEEE/ASME Transactions on Mechatronics}, 22(2),  717--727.

\bibitem{wang}
Wang, J., Fei, Y. (2019)
Design and Modelling of Flex-Rigid Soft Robot for Flipping Locomotion. \emph{J. Intell. Robot. Syst.}, doi: 10.1007/s10846-018-0957-7.

\bibitem{zheng}
Zheng, K., Hu, Y., Wu, B., Guo, X. (2019) New trajectory control method for robot with flexible bar-groups based on workspace lattices.
\emph{Robotics Aut. Syst.}, 111, 44--61.

\end{thebibliography}
\end{document}